\documentclass{JHEP3}
\usepackage{graphics}
\usepackage{graphicx}
\usepackage{amsmath}
\usepackage{cite}
\newcommand{\PL}[2]{\, {\rm Li}_{#1}\!\left({#2}\right)}

\newcommand{\ESGamma}{S_{\Gamma}}

\def\be{\begin{equation}}
\def\ee{\end{equation}}
\def\bea{\begin{eqnarray}}
\def\eea{\end{eqnarray}}
\def\eps{\epsilon}
\def\nnb{\nonumber}
\def\eps{\epsilon}
\def\dps{\displaystyle}

\def\cG{  {\cal G}  }
\def\cN{  {\cal N}  }
\def\cO{  {\cal O}  }
\def\Tr{  {\rm Tr}  }

\def\cA {  {\cal A}  }

\preprint{HU-EP-11/11/61\\
NSF-KITP-11-268\\
ZU-TH 28/11\\
SI-HEP-2011-19}
\title{The three-loop form factor in $\mathcal{N}=4$ super Yang-Mills}

\author{Thomas  Gehrmann$^{a,b}$, Johannes M.\ Henn$^{b,c,d}$, Tobias Huber$^{b,e}$\\
$^a$ Institut f\"ur Theoretische Physik, Universit\"at Z\"urich, \\
Winterthurerstrasse 190, CH-8057 Z\"urich, Switzerland\\
$^b$ Kavli Institute for Theoretical Physics \\
University of California, Santa Barbara,  CA 93106, USA \\
$^c$Institut f\"ur Physik, Humboldt-Universit\"at zu Berlin, \\
Newtonstra\ss{}e15, D-12489 Berlin, Germany \\
$^d$ Institute for Advanced Study, Princeton, NJ 08540, USA\\
$^e$ Theoretische Physik 1, Naturwissenschaftlich-Technische Fakult\"at,
\\ Universit\"at Siegen, Walter-Flex-Strasse 3, D-57068 Siegen, Germany\\

\email{thomas.gehrmann@uzh.ch, \\ jmhenn@ias.edu, \\ huber@tp1.physik.uni-siegen.de}}

\abstract{
In this paper we study the Sudakov form factor in $\cN=4$ super Yang-Mills 
theory to the three-loop order.
The latter is expressed in terms of planar and non-planar loop integrals. 
We show that it is possible to choose a representation in which each loop integral has 
uniform transcendentality.
We verify analytically the expected exponentiation of the infrared divergences with the 
correct values of the three-loop cusp and collinear anomalous dimensions in dimensional
regularisation.
We find that the form factor in $\cN=4$ super Yang-Mills can be related
to the leading transcendentality part of the quark and gluon form factors in QCD. 
We also study the ultraviolet properties of the form factor in $D>4$ dimensions,
and find unexpected cancellations, resulting in an improved ultraviolet behaviour.
}


\begin{document}

\section{Introduction}\label{sec:intro}

In this paper we study the Sudakov form factor in $\cN=4$ super Yang-Mills (SYM) with gauge group $SU(N)$. Following van Neerven \cite{vanNeerven:1985ja}, we study the vacuum expectation value 
of an operator built from two scalars, inserted into two on-shell states. The operator belongs to the stress-energy supermultiplet, which contains the conserved currents of $\cN=4$ SYM, and 
has zero anomalous dimension. Together with the vanishing $\beta$ function of 
$\cN=4$ SYM this means that the form factor is ultraviolet (UV) finite in four dimensions. 
Therefore only infrared (IR) divergences associated to the on-shell states appear, which we 
regularise using dimensional regularisation.

Generalisations of the Sudakov form factor to the case of 
different composite operators, and more external on-shell legs, have been discussed recently in refs. \cite{Brandhuber:2010ad,Brandhuber:2011tv,Bork:2010wf,Bork:2011cj}. 
Form factors have also been studied
within the AdS/CFT correspondence in the dual AdS 
description, see refs. \cite{arXiv:0710.1060,Maldacena:2010kp}.
Here we will focus on the perturbative expansion of the form factor of ref. \cite{vanNeerven:1985ja}.

Form factors are closely related to scattering amplitudes. For example, planar amplitudes
can be factorised into an infrared divergent part, given by a product of
form factors, and an infrared finite remainder (a `hard' function in QCD
terminology), see e.g.\ ref.~\cite{Bern:2005iz} and references therein.
The infrared divergent part exponentiates and has a simple universal form.
In fact, for four- and five-point scattering amplitudes the exponentiation property of the 
divergent part carries over to the finite part as well~\cite{Bern:2005iz,Anastasiou:2003kj}. This is a consequence of
a hidden dual conformal symmetry of planar scattering amplitudes. 
The latter relates the finite part to the infrared divergent part through a Ward identity 
\cite{Drummond:2007cf,Drummond:2007au}.
The relation to form factors makes it possible to give an operator definition of the finite remainder. The scheme independence of the latter was recently checked in a two-loop computation
using dimensional and massive regularisations \cite{Henn:2011by}.

Scattering amplitudes in $\cN=4$ SYM have many special properties, and
it is interesting to ask how much of this simplicity carries over to the form factors.
For both the planar four-particle amplitude and the form factor, the general form of the 
result is known in principle. For the former, this is due to dual conformal symmetry, and
for the latter it is due to the exponentiation of infrared divergences.
However it is quite non-trivial to obtain these a priori known results from explicit
perturbative calculations, evaluating loop integrals. The simplicity of 
the final results suggests that there should be more structure hidden in
the loop integral expressions, and by studying them further one might
gain insights into better ways of evaluating them, which is of more general
interest.

One might expect that the evaluation of form factors should be simpler than that of 
scattering amplitudes, as the former have a trivial scale dependence only, 
whereas the latter are functions of ratios of Mandelstam variables, e.g.\ $s/t$
in the four-point case.
Given this, it is somewhat surprising that less is known about the loop expansion
of form factors in $\cN=4$ SYM than about scattering amplitudes. 
For example, while the planar four-point amplitude was evaluated to
the four-loop order (in part numerically) \cite{hep-th/0610248,hep-th/0612309,arXiv:1004.5381}, 
the form factor has only been computed to the
two-loop order in ref. \cite{vanNeerven:1985ja}, in a calculation that dates back to 1986.
In the present paper, we extend the calculation of ref. \cite{vanNeerven:1985ja} to three loops, 
and study which of the properties that have been observed for scattering amplitudes are 
present.

One fact which makes form factors technically challenging compared to planar amplitudes,
however also more interesting, is the following. At leading order in the `t~Hooft limit
$N \to \infty$,  where the coupling $\lambda = g^2 N$ is kept fixed, both planar as well as
non-planar integrals appear in the form factor.
This is easily understood by the fact that the operator insertion is a colour-singlet. 
It is interesting to note that the non-planar diagrams appearing in the 
 form factor are related, through the unitarity technique, to a priori subleading 
 double trace terms in the four-particle scattering amplitude.
Therefore, the form factor at leading order in $N$ contains information about
non-planar corrections to the four-particle amplitude. 
The first non-planar diagram, the crossed ladder, appears at the two-loop level. 
At three loops, we find five different non-planar diagrams that contribute, i.e.\ that
have non-vanishing coefficient.

It is an observed, albeit unproven fact that results for scattering amplitudes in $\cN=4$ super 
Yang-Mills have uniform transcendentality (UT), i.e.\ can be expressed as linear combinations
of polylogarithmic functions of uniform degree $2L$, where $L$ is the loop order, with 
constant coefficients. In $\eps$-expansions of dimensionally regularised quantities 
which depend only on a single scale, the coefficients of the  
Laurent expansion in $\eps$ are real constants 
which are in general of increasing transcendentality in the Riemann
$\zeta$-function. In this context uniform transcendentality refers to
homogeneity in the degree of transcendentality ($DT$), where the latter is 
defined as
\bea
DT(r)&=&0\quad\mbox{ for rational } r\nonumber\\
DT(\pi^k)&=&DT(\zeta_k)=k\nonumber\\
DT(x\cdot y)&=&DT(x)+DT(y) \nonumber \;.
\eea
In the planar case, the property of UT is even true for individual loop integrals, 
at least when they are expressed in an appropriate basis of dual conformal integrals
\cite{ArkaniHamed:2010gh,Drummond:2010mb}. Incidentally, this also has practical 
advantages, as these integrals are
easier to evaluate \cite{Drummond:2010mb,arXiv:1010.3679} than those in other representations.
Dual conformal symmetry is only expected in the planar case,
but what can be said about the transcendentality properties of non-planar integrals?
At four points, the non-planar double ladder integral is not of uniform transcendentality. 
However, if defined with an appropriate loop-dependent numerator factor, 
it does have this property \cite{Lance,hep-ph/9909506}. 
Changing to a basis involving the latter integral allows one to understand the UT property of four-point 
non-planar $\cN=4$ SYM amplitudes \cite{Naculich:2008ys} and $\cN=8$ supergravity amplitudes \cite{Naculich:2008ew,Brandhuber:2008tf}.
It also raises the interesting question whether this is a generic feature.

All planar and non-planar master integrals for form 
factors in dimensional regularisation at three loops are known from the
computation of the form factor in
QCD~\cite{masterA,masterB,masterC,BCSSS,arXiv:1001.2887,arXiv:1001.3132,Gehrmann:2010ue,arXiv:1005.0362,Lee:2010ik,arXiv:1010.4478},
and some of them have UT, while others do not. It has been observed that some of the integrals do have UT if they are defined with
certain (loop-dependent) numerator factors \cite{Lance}. The latter resemble the numerator 
factors required by dual conformal symmetry in the planar case \cite{hep-th/0607160}. In this paper, we find
similar numerator factors for all topologies with $7,8,9$ propagators, such that the integrals have UT.
Moreover, we find that  the complete three-loop form factor can be written solely in terms of UT integrals.

Finding a representation that has this property required using certain identities 
for non-planar form factor integrals that are based on reparametrisation invariances,
which we found as a by-product of our analysis.
They generalise an identity found by Davydychev and Usyukina \cite{hep-ph/9412356}.

As was already mentioned, in $\cN=4$ SYM, scattering amplitudes and the 
form factors studied here are UV finite in four dimensions. 
It is interesting to ask in what dimension, called critical dimension $D_{c}$, 
they first develop UV divergences. 
This question is of theoretical interest in the context of the discussion of possible
finiteness of $\cN=8$ supergravity, see e.g.\ \cite{Bern:2007hh,Stelle:2007zz,Bern:2009kd} and 
references therein. More practically,
bounds on the critical UV dimension at a given loop order
can also be a useful cross-check of computations, or constrain the types of loop
integrals that can appear. Ultraviolet power counting, based on the existence
of $\cN=3$ off-shell superspace \cite{570842}, provides a lower bound for the critical dimension.
We analyse the UV properties of the form factor to three loops and find that at each loop
order, the critical dimension is $D_c=6$. This is consistent with the bound obtained from
superspace power counting. 
We find that the latter bound is saturated at two loops, while it is too conservative at
three loops, where the ultraviolet behaviour is better than suggested by the bound.
This is the result of a cancellation between different loop integrals.
We find a representation where the UV behaviour is manifest.

This paper is organised as follows. We review the known expression for the form factor
to two loops in section \ref{sec:ff2l}. We then discuss identities for non-planar
integrals to three loops in section \ref{sec:rpi}. In section \ref{sec:ff3luni}, using 
the unitarity-based method, we derive an expression
for the three-loop form factor in terms of loop integrals. 
We then evaluate the latter in section \ref{sec:ff3l} and verify the exponentiation
of infrared divergences in section \ref{sec:logff}.
We then analyse the ultraviolet properties of the form factor to three loops in section \ref{sec:uv}.
We conclude in section \ref{sec:conc}.
There are several appendices. Appendix~\ref{app:integrals} contains the analytic expressions of the 
$\epsilon$ expansion of the 
integrals used in the paper, while appendix~\ref{app:ffmasters} contains 
the expression of the form factor in terms of conventionally
used master integrals. Finally, appendix~\ref{app:4pt} reviews the on-shell four-point amplitude
to two loops that is used in the unitarity calculation in the main text.

\tableofcontents


\section{Form factor to two loops}\label{sec:ff2l}

In order to define the scalar form factor in $\mathcal{N}=4$
SYM, we start by introducing the bilinear operator
\be\label{eq:composite}
\cO = \Tr(\phi_{12} \phi_{12}) \,,
\ee 
where the scalars $\phi_{AB}$ are in the representation $\bf 6$ of $SU(4)$,
and $\phi_{AB} = \phi_{AB}^{a} T_{a}$, with $T_{a}$ being
the generators of $SU(N)$ in the fundamental representation,
normalised according to $\Tr(T^{a} T^{b})=\delta^{ab}$.
This operator is a particular component of the stress-energy supermultiplet
of $\cN=4$ SYM, and has zero anomalous dimension.
We then define the form factor as the vacuum expectation value of
$\cO $ inserted into two on-shell states in the adjoint representation,
\be
\mathcal{F}_{\cal S} = \langle \phi^{a}_{34}(p_1) \phi^{b}_{34}(p_2)\, \cO \rangle \,,
\ee
with the convention that momentum is outgoing.

Since $\cO$ is a colour singlet, the form factor must be proportional to $ \Tr(T^{a} T^{b})$,
\be\label{eq:definitionFS}
\mathcal{F}_{\cal S} = \Tr(T^{a} T^{b}) F_{\cal S} \,.
\ee
We work in dimensional regularisation with $D=4-2\eps$
dimensions in order to regulate IR divergences associated with the on-shell legs.
We write the form factor as an expansion in the
't~Hooft coupling~\cite{Bern:2005iz}
\be
\dps a = \frac{g^2 \, N}{8\pi^2} \,
(4\pi)^\eps \, e^{-\eps\gamma_E}\,,
\ee
according to
\bea\label{eq:ffexpansion1}
&& \nnb \\
\dps {F}_S &=& 1+ a \, x^{\eps} \, {F}_S^{(1)}
+ a^2 \, x^{2\eps} \, {F}_S^{(2)}
+ a^3 \, x^{3\eps} \, {F}_S^{(3)}+{\cal{O}}(a^4)\; . \\
&& \nnb
\eea
We normalized the tree-level contribution to unity and
introduced
\be
\dps x = \frac{\mu^2}{-q^2-i\eta}  \; ,
\ee
with the infinitesimal quantity $\eta>0$.

We remark that the dependence on the number of colours $N$ in equation (\ref{eq:ffexpansion1}) is 
exact. In order to see this, let us show that the three-loop contribution to the form factor must be 
proportional to $N^3$ (a similar analysis trivially holds at one and two loops).

The reasoning is as follows. Imagine a generic Feynman diagram contributing to $\mathcal{F}_{\cal S}$.
Without loss of generality, suppose that it is built from three-point vertices, whose colour dependence is given
by the structure constants $f^{a_1 a_2 a_3}$. For each internal line, there is a sum over adjoint colour indices, with
the result being proportional to $\Tr(T^{a} T^{b})$, as stated in equation (\ref{eq:definitionFS}). Our goal
is to determine the proportionality factor. In order to do this, it is convenient to sum also over the free indices $a$ and $b$,
\be \label{eq:extrafactor}
\sum_{a,b} \delta^{ab} {\rm Tr}( T^a T^b ) = N^2 -1 \,.
\ee 
We can then represent each Feynman diagram as a circle with inscribed lines.
There are three inequivalent structures that can appear, 
\begin{eqnarray}
A&=&f^{abg} f^{bcg}f^{c d h} f^{e  d i} f^{e  f i} f^{f a h}\,,  \nonumber \\
B&=&f^{a b g} f^{b c h} f^{c d g} f^{d e i} f^{e i f} f^{f h a} \,, \\
C&=&f^{a b g} f^{b c h} f^{c d i} f^{d e g} f^{e h f} f^{f i a}  \,, \nonumber
\end{eqnarray}
which correspond to the case of zero, one, or two intersections of the
inscribed lines, respectively. Sums over repeated indices are implicit.
In order to carry out the sums, it is convenient to write the structure constants as 
\be
f^{abc} = -i/\sqrt{2} \left({\rm Tr}[T^a T^b T^c] - {\rm Tr}[T^a T^c T^b]\right)\,.
\ee
Using the $SU(N)$ Fierz identities,
\bea
\sum_{a} \Tr( A T^{a} ) \Tr(B T^{a} ) &=&  \Tr(A B) - 1/N \, \Tr(A) \, \Tr(B) \,,  \label{Fierz1}\\
\sum_{a} \Tr( A T^{a}  B T^{a}  ) &=&  \Tr(A) \, \Tr(B) - 1/N  \, \Tr(AB) \,, \label{Fierz2}
\eea
one easily finds
\be
A = (N^2 - 1) N^3\,,\qquad B= -\frac{1}{2} (N^2 - 1) N^3 \,,\qquad C=0\,.
\ee
Taking into account equation (\ref{eq:extrafactor}), we see that $F_{S}$ at three loops is proportional to $N^3$, as claimed.

Note that beginning from four loops there can be more than one colour structure,
and in particular the quartic Casimir can appear. An explicit example of this is the 
four-loop contribution to the QCD $\beta$ function \cite{vanRitbergen:1997va}. 
An interesting related question has to do with the colour 
dependence of infrared divergences in gauge theories, see e.g. \cite{Becher:2009qa},
and references therein.

\FIGURE[t]{
\includegraphics[width=0.95\textwidth]{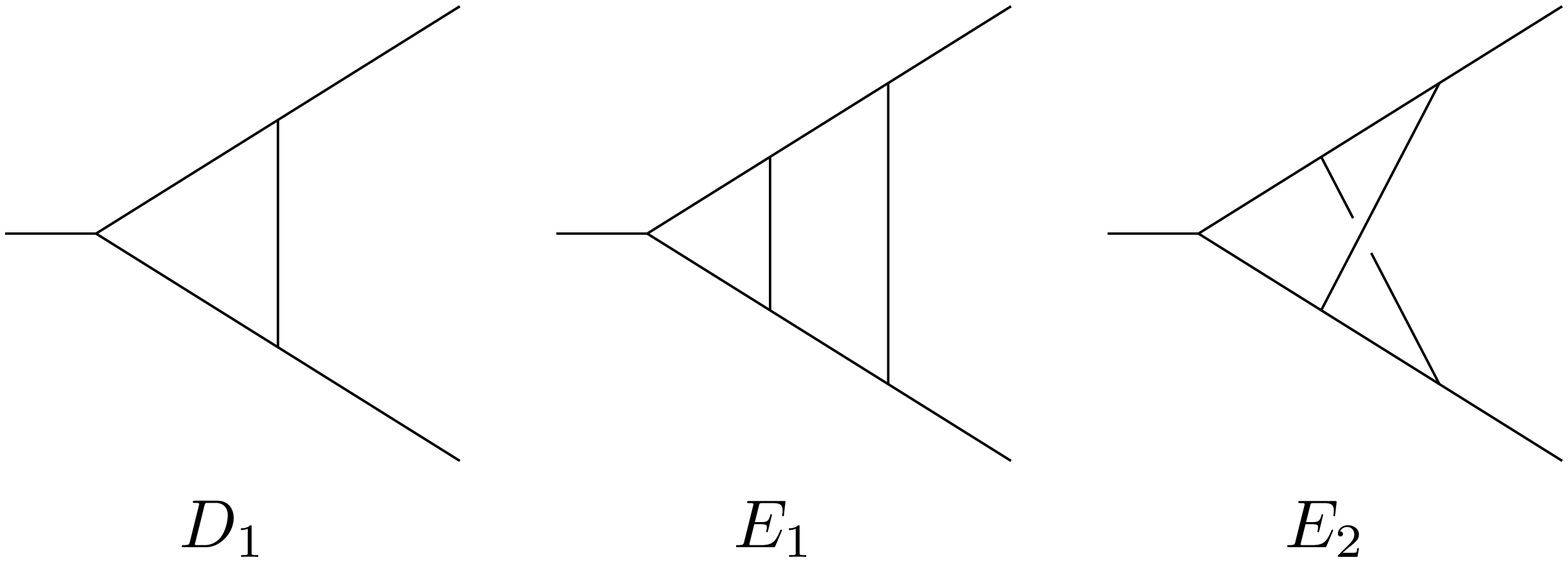}
\caption{Diagrams that contribute to the one-loop and
two-loop form factors $\mathcal{F}_S^{(1)}$ and
$\mathcal{F}_S^{(2)}$ in $\mathcal{N}=4$ SYM. All internal
lines are massless. The incoming momentum is $q=p_1+p_2$,
outgoing lines are massless and on-shell, i.\ e.\ $p_1^2 =
p_2^2 = 0$. All diagrams displayed have unit numerator and
exhibit uniform transcendentality (UT) in their Laurent
expansion in $\eps = (4-D)/2$.}
\label{fig:diagsDE}}

The form factor to two loops was computed a long
time ago~\cite{vanNeerven:1985ja}. It contains as building
blocks the diagrams displayed in Fig.~\ref{fig:diagsDE} and
reads
\bea
{F}_S &=& 1+ g^2 \, N \, \mu^{2\eps} \cdot (-q^2)\cdot 2\, D_1
+ g^4 \, N^2 \, \mu^{4\eps} \cdot (-q^2)^2 \cdot
\left[4 \, E_1 + E_2\right] +{\cal{O}}(g^6) \nnb \\
&& \nnb \\
&=& 1+ a \, x^{\eps} \, R_{\eps} \cdot 2 \, D_1^{{\rm exp}}
+ a^2 \, x^{2\eps} \, R_{\eps}^2
\cdot \left[4 \, E_1^{{\rm exp}} + E_2^{{\rm exp}}\right]
+{\cal{O}}(a^3)\; , \label{eq:ffexpansion}
\eea
with
\be\label{eq:Reps}
\dps R_{\eps} \equiv \frac{e^{\eps\gamma_E}}{2 \, \Gamma(1-\eps)} \; .
\ee
The expressions for $D_1$, $D_1^{{\rm exp}}$, $E_i$, and $E_i^{{\rm
exp}}$ are given explicitly in appendix~\ref{app:integrals} and result
in
{\allowdisplaybreaks
\bea
{F}_S^{(1)} &=& R_{\eps} \cdot 2 \, D_1^{{\rm exp}} \nnb \\
&=& -\frac{1}{\eps^2}+\frac{\pi ^2}{12}+\frac{7\zeta_3}{3} \, \eps
+ \frac{47 \pi ^4}{1440} \, \eps^2 +\eps^3 \left(\frac{31
\zeta_5}{5}-\frac{7 \pi ^2\zeta_3}{36}\right) +\eps^4 \left(\frac{949
\pi ^6}{120960}-\frac{49\zeta_3^2}{18}\right)\nnb \\
&& +\eps^5\left(-\frac{329
\pi ^4 \zeta_3}{4320}-\frac{31 \pi ^2 \zeta_5}{60}+\frac{127
\zeta_7}{7}\right) +\eps^6 \left(\frac{49 \pi ^2
\zeta_3^2}{216}-\frac{217 \zeta_3 \zeta_5}{15}+\frac{18593 \pi
^8}{9676800}\right) \nnb \\
&& + {\cal{O}}(\eps^7) \; , \label{eq:1lff} \\
&& \nnb \\
{F}_S^{(2)} &=& R_{\eps}^2
\cdot \left[4 \, E_1^{{\rm exp}} + E_2^{{\rm exp}}\right] \nnb \\
&=&+\frac{1}{2 \eps^4}-\frac{\pi ^2}{24\eps^2}-\frac{25 \zeta_3}{12
\eps}-\frac{7 \pi^4}{240}
+\eps \left(\frac{23 \pi ^2 \zeta_3}{72}+\frac{71 \zeta_5}{20}\right)
+\eps^2 \left(\frac{901 \zeta_3^2}{36}+\frac{257 \pi
^6}{6720}\right)\nnb \\
&&+\eps^3 \left(\frac{1291 \pi
   ^4 \zeta_3}{1440}-\frac{313 \pi ^2 \zeta_5}{120}+\frac{3169
   \zeta_7}{14}\right) \nnb \\
&&+\eps^4 \left(-66 \zeta_{5,3}+\frac{845 \zeta_3
\zeta_5}{6}-\frac{1547 \pi ^2 \zeta_3^2}{216}+\frac{50419 \pi
   ^8}{518400}\right)+ {\cal{O}}(\eps^5) \; . \label{eq:2lff}
\eea
}
The multiple zeta values $\zeta_{m_1,\dots,m_k}$ are defined by (see
e.g.~\cite{Blumlein:2009cf} and references therein)
\be\label{eq:MZVdef}
\zeta_{m_1,\dots,m_k}=\sum\limits_{i_1=1}^\infty\sum\limits_{i_2=1}^{i_1-1}
\dots\sum\limits_{i_k=1}^{i_{k-1}-1}
\prod\limits_{j=1}^k\frac{\mbox{sgn}(m_j)^{i_j}}{i_j^{|m_j|}}\,.
\ee
The numerical values of the transcendental constants up to weight eight are:
\begin{eqnarray*}
&& \zeta_3 = 1.2020569031595942854\ldots\,, \qquad 
\zeta_5= 1.0369277551433699263\ldots\, ,\\
&& \zeta_7 = 1.0083492773819228268\ldots\,,\qquad 
\zeta_{5,3} = 0.037707672984847544011\ldots\, .
\end{eqnarray*}


\section{Momentum routing invariances of integrals}\label{sec:rpi}

Before we proceed to calculate the $\cN=4$ SYM form factor to three
loops via unitarity cuts, we want to investigate some of the occurring
topologies more closely. In particular, we will derive identities that
relate integrals without uniform transcendentality (UT) to integrals
that do have this property. Since the diagrams that we will obtain from
the unitarity method do not individually have UT, the following relations
will be very useful later on for switching to an integral basis
for the form factor in which each building block has UT.

We start with topology $F_3^{\ast}$, see Fig.~\ref{fig:diagsRPI1}. We
label its incoming momentum with $q=p_1+p_2$, and the outgoing ones
with $p_1$ and $p_2$, respectively. The latter are massless and
on-shell, i.e.\ $p_1^2 = p_2^2 = 0$. The topology can be parametrised
according to
\be\label{eq:linesF3star}
\left\{ k_1 - k_2 \, , \, k_1 - k_3 \, , \, k_1 - k_2 - k_3 \, ,
 \, k_2 \, , \, k_3 \, , \, k_1 - q \, ,
 \, k_2 - q \, , \, k_3 - q \, , \, k_2 - p_1 \right\} \; , 
\ee
where $k_i$ are the loop momenta. It can be seen from
Fig.~\ref{fig:diagsRPI1} how the momenta are distributed among the lines
of the diagram $F_3^{\ast}$. It turns out that the following 
reparametrization of loop momenta,
{\allowdisplaybreaks
\bea
k_1 &\to& q+k_2-k_1 \nnb \\
k_2 &\to& k_2 \nnb \\
k_3 &\to& q-k_3 \; , \nnb
\eea
}
does not only leave the value of the integral invariant, but even its
{\textit{integrand}}. We can now apply this transformation to the
integral $F_3$ which carries the factor $(k_2-k_3)^2$ as an
irreducible scalar product in its numerator. This yields
{\allowdisplaybreaks
\bea
(k_2-k_3)^2 &\to& (k_2 + k_3 - q)^2 \nnb \\
    && = k_2^2 + k_3^2 + (k_2 - q)^2 + (k_3 - q)^2
         - (k_2-k_3)^2 - q^2 \; .
\eea
}
We can now solve this equation for $(k_2-k_3)^2$ and get the following
relation between integrals,
\be\label{eq:RPIF3}
\dps F_3 = -\frac{1}{2} \, q^2 \, F_3^{\ast} + F_{a1} + F_8 \; ,
\ee
which is diagrammatically shown in Fig.~\ref{fig:diagsRPI1}. We have
now decomposed the integral $F_3^{\ast}$, which does not have
UT in its Laurent expansion, into two
integrals ($F_3$ and $F_8$) which indeed do have this property, and
the auxiliary integral $F_{a1}$, which again does not have homogeneous
transcendental weight, but which will be cancelled later on.

\FIGURE[t]{
\includegraphics[width=0.95\textwidth]{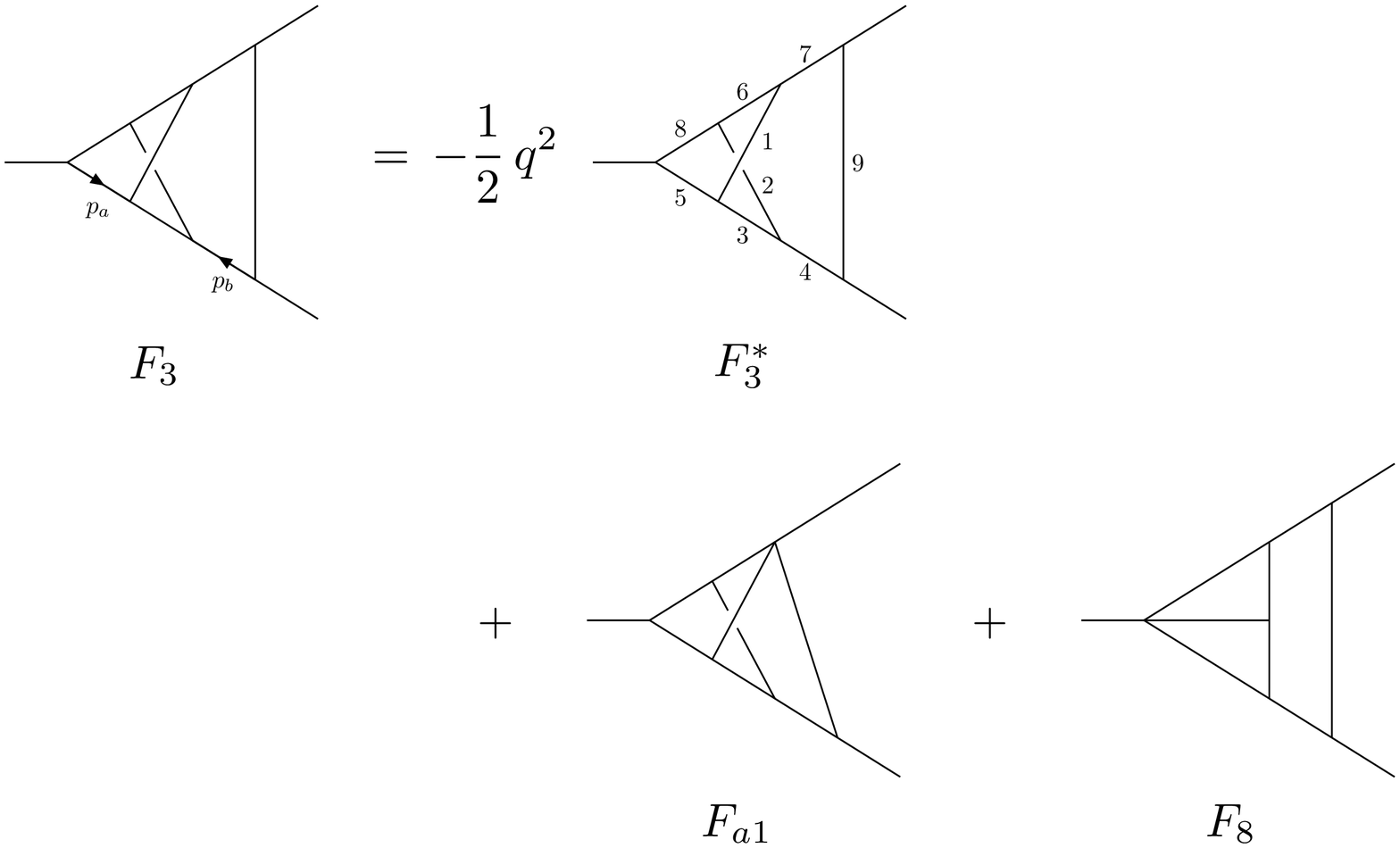}
\caption{Diagrammatic representation of Eq.~(\ref{eq:RPIF3}).
The internal lines of all diagrams are massless. The incoming momentum is
$q=p_1+p_2$, outgoing lines are massless and on-shell, i.e.\ $p_1^2 = p_2^2 = 0$.
Diagrams with labels $p_a$ and $p_b$ on arrow lines have an irreducible scalar product
$(p_a+p_b)^2$ in their numerator (diagrams that lack these
labels have unit numerator). The numbers in $F_3^{\ast}$ indicate the
position of the entries in Eq.~(\ref{eq:linesF3star}).
Diagrams $F_3$ and $F_8$ have UT, contrary to $F_3^{\ast}$ and $F_{a1}$.}
\label{fig:diagsRPI1}}

We can apply analogous steps to topology $F_4^{\ast}$, see
Fig.~\ref{fig:diagsRPI2}. The topology can be parametrised
according to
\be\label{eq:linesF4star}
\left\{ k_1 \, , \, k_2 \, , \, k_3 \, ,
 \, k_1 - k_2 \, , \, k_1 - k_3 \, , \, k_1 - q \, ,
 \, k_1 - k_2 - p_2 \, , \, k_3 - q \, , \, k_2 - p_1 \right\} \; , 
\ee
and the distribution of the momenta among the lines
can be seen from Fig.~\ref{fig:diagsRPI2}.
The integrand remains invariant under
{\allowdisplaybreaks
\bea
k_1 &\to& q-k_1 \nnb \\
k_2 &\to& p_1 - k_2 \nnb \\
k_3 &\to& q-k_3 \; , \nnb
\eea
}
We now apply this transformation to the numerator $(k_1-p_1)^2$ of the
integral $F_4$. This yields
{\allowdisplaybreaks
\bea
(k_1-p_1)^2 &\to& (k_1 - p_2)^2 \nnb \\
    && = k_1^2 + (k_1 - q)^2 - (k_1-p_1)^2 - q^2 \; .
\eea
}
\FIGURE[t]{
\includegraphics[width=0.95\textwidth]{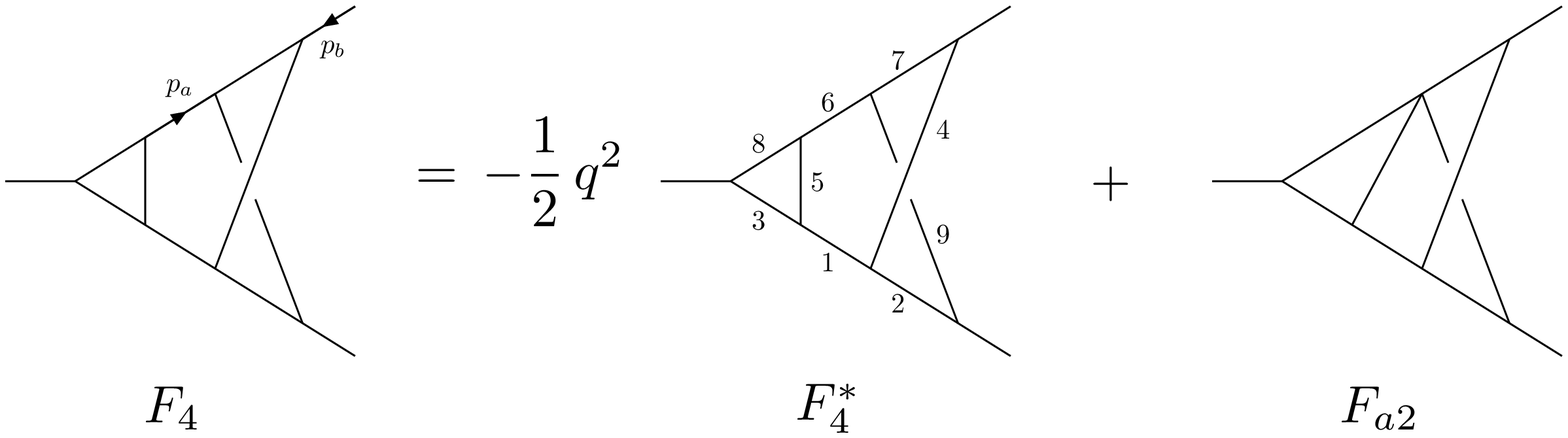}
\caption{Diagrammatic representation of Eq.~(\ref{eq:RPIF4}).
All symbols have the same meaning as in Fig.~\ref{fig:diagsRPI1}.
The numbers in $F_4^{\ast}$ indicate the
position of the entries in Eq.~(\ref{eq:linesF4star}).
Diagram $F_4$ has UT, contrary to $F_4^{\ast}$ and $F_{a2}$.}
\label{fig:diagsRPI2}}
We can now solve this equation for $(k_1-p_1)^2$ and get
\be\label{eq:RPIF4}
\dps F_4 = -\frac{1}{2} \, q^2 \, F_4^{\ast} + F_{a2} \; ,
\ee
which is diagrammatically shown in Fig.~\ref{fig:diagsRPI2}. Again we
decomposed the non-homogeneous integral $F_4^{\ast}$ into the
homogeneous integral $F_4$ and yet another non-homogeneous
auxiliary integral ($F_{a2}$) which will be cancelled later on.

We can also decompose the topology $F_5^{\ast}$, see
Fig.~\ref{fig:diagsRPI3}. In this case we cannot find a relation
between integrals which is based on a momentum routing invariance,
but a relation which is simply based on momentum conservation. The
topology can be parametrised according to
\be\label{eq:linesF5star}
\left\{ k_1 - k_2 \, , \, k_1 - k_3 \, , \, k_1 - k_2 - k_3 \, ,
 \, k_2 \, , \, k_3 \, , \, k_1 - q \, ,
 \, k_2 - q \, , \, k_1 - p_1 \, , \, k_3 - p_1 \right\} \; ,
\ee
and we refer to Fig.~\ref{fig:diagsRPI3} for their distributions
among the lines. From momentum conservation we get
\be
(k_2 - p_1)^2 = (k_1-k_2)^2 - k_1^2 + k_2^2 + (k_1 - p_1)^2 -
(k_1-k_2-p_1)^2 \; ,
\ee
which results in
\be\label{eq:RPIF5}
\dps F_5^{\ast} = F_{a1} + F_{a2} + F_9 - F_5 - F_6 \; .
\ee
Hence we decomposed $F_5^{\ast}$ into the homogeneous-weight diagrams
$F_5$, $F_6$, and $F_9$, as well as the same non-homogeneous diagrams
$F_{a1}$, and $F_{a2}$ which already appeared above.
\FIGURE[t]{
\includegraphics[width=0.95\textwidth]{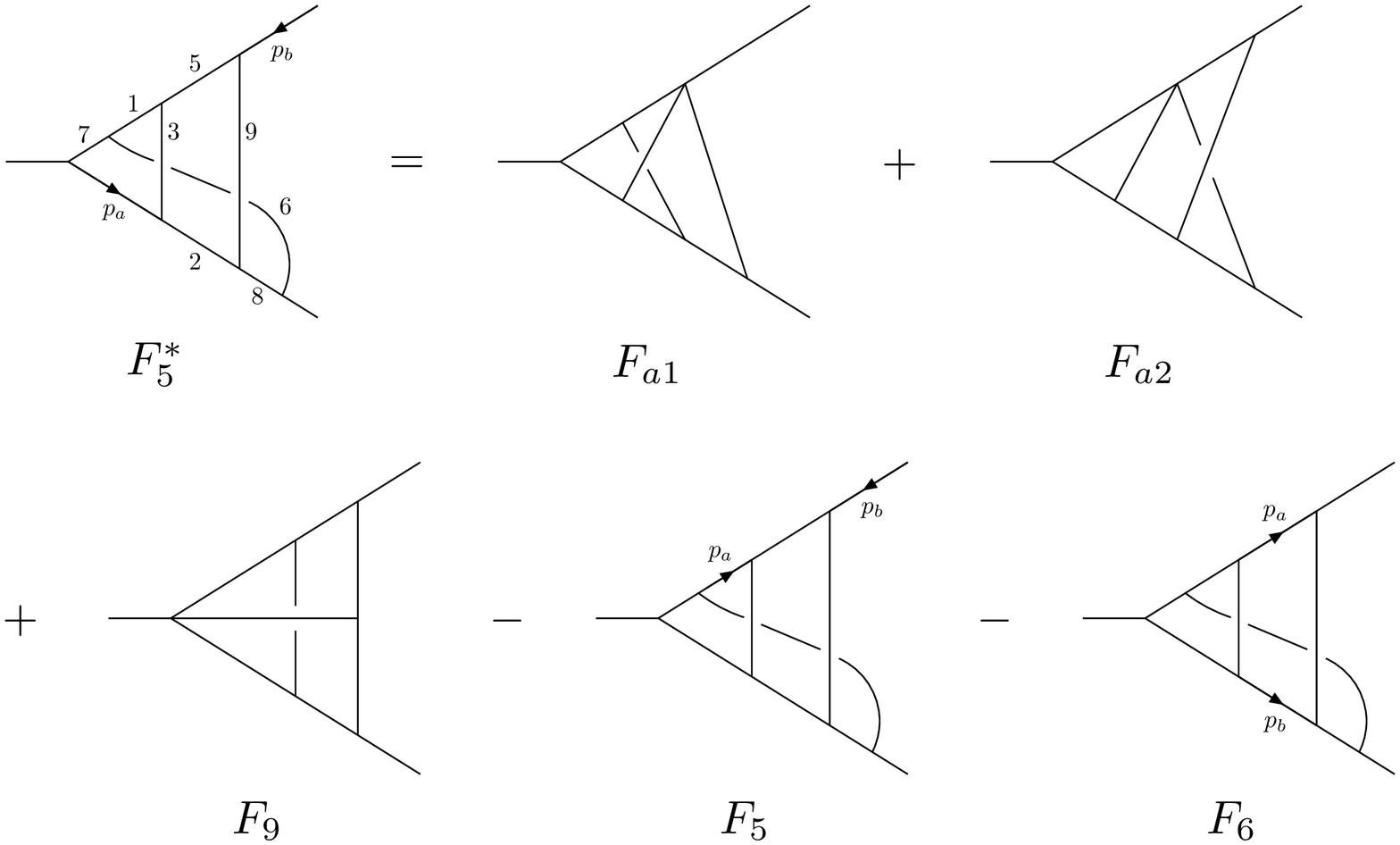}
\caption{Diagrammatic representation of Eq.~(\ref{eq:RPIF5}).
All symbols have the same meaning as in Fig.~\ref{fig:diagsRPI1}.
The numbers in $F_5^{\ast}$ indicate the
position of the entries in Eq.~(\ref{eq:linesF5star}).
Diagrams $F_5$, $F_6$, and $F_9$ have UT, contrary to $F_5^{\ast}$, $F_{a1}$, and $F_{a2}$.}
\label{fig:diagsRPI3}}

We see from Eqs.~(\ref{eq:RPIF3}),~(\ref{eq:RPIF4}),
and~(\ref{eq:RPIF5}) that only two auxiliary diagrams of
non-homogeneous weight, namely $F_{a1}$, and $F_{a2}$ appear in all
these relations. It turns out that the coefficients obtained from
unitarity are precisely such that these integrals cancel in the
expression for the form factor.

We checked all relations between integrals also at the level of their
integration-by-parts (IBP) 
reduction~\cite{chet}
to master integrals using the 
implementation of the Laporta algorithm~\cite{Laporta:2001dd} in the REDUZE~\cite{Studerus:2009ye} 
code. We find that all  relations obtained from momentum routing invariance 
in this section can actually be reproduced from solving IBP relations, which 
is a priori not guaranteed for a general Feynman integral topology.  
 The $\eps$-expansions of all integrals can be found in
appendix~\ref{app:integrals}.


\section{Form factor to three loops from unitarity cuts}\label{sec:ff3luni}

Here we use unitarity cuts to derive an expression for the three-loop form factor in terms
of the integrals discussed in the previous section. 
We will compute the form factor in a perturbative expansion in the Yang-Mills coupling $g$,
and denote the contribution at order $g^0, g^2, g^4, g^6$ by ${\cal F}_{S}^{tree}, {\cal F}_{S}^{1-loop}, 
{\cal F}_{S}^{2-loop}, {\cal F}_{S}^{3-loop}$, respectively, and similarly for $F_{S}$. 
Note that this notation, convenient for the unitarity calculations, differs from the one used in Eq.~(\ref{eq:ffexpansion1}).

The essential features of the unitarity-based method \cite{Bern:1994zx,Britto:2004nc}
that we are going
to use are reviewed in the recent paper \cite{Carrasco:2011hw}.
We will employ two-particle cuts, as well as generalised cuts. 
The two-particle cuts are very easy to evaluate,
and we show an explicit example below. 

In order to evaluate more complicated cuts, with many intermediate state
sums to be carried out, it is extremely useful to employ
a formalism that makes supersymmetry manifest. This can be done by 
arranging the on-shell states of $\cN=4$ SYM into an on-shell supermultiplet
\cite{Nair:1988bq}.
The main advantage is that intermediate state sums appearing in the cuts
become simple Grassmann integrals that can be carried out trivially 
\cite{Drummond:2008bq,Bern:2009xq,Elvang:2008na}. In this way, it is easy to obain
compact analytical expressions for the cuts.

We follow the notations for unitarity cuts of ref. \cite{hep-ph/9702424}.
We start by reviewing the one- and two-loop cases as examples.

\subsection{One-loop form factor from unitarity cuts}

{}
\FIGURE[t]{
\includegraphics[width=0.99\textwidth]{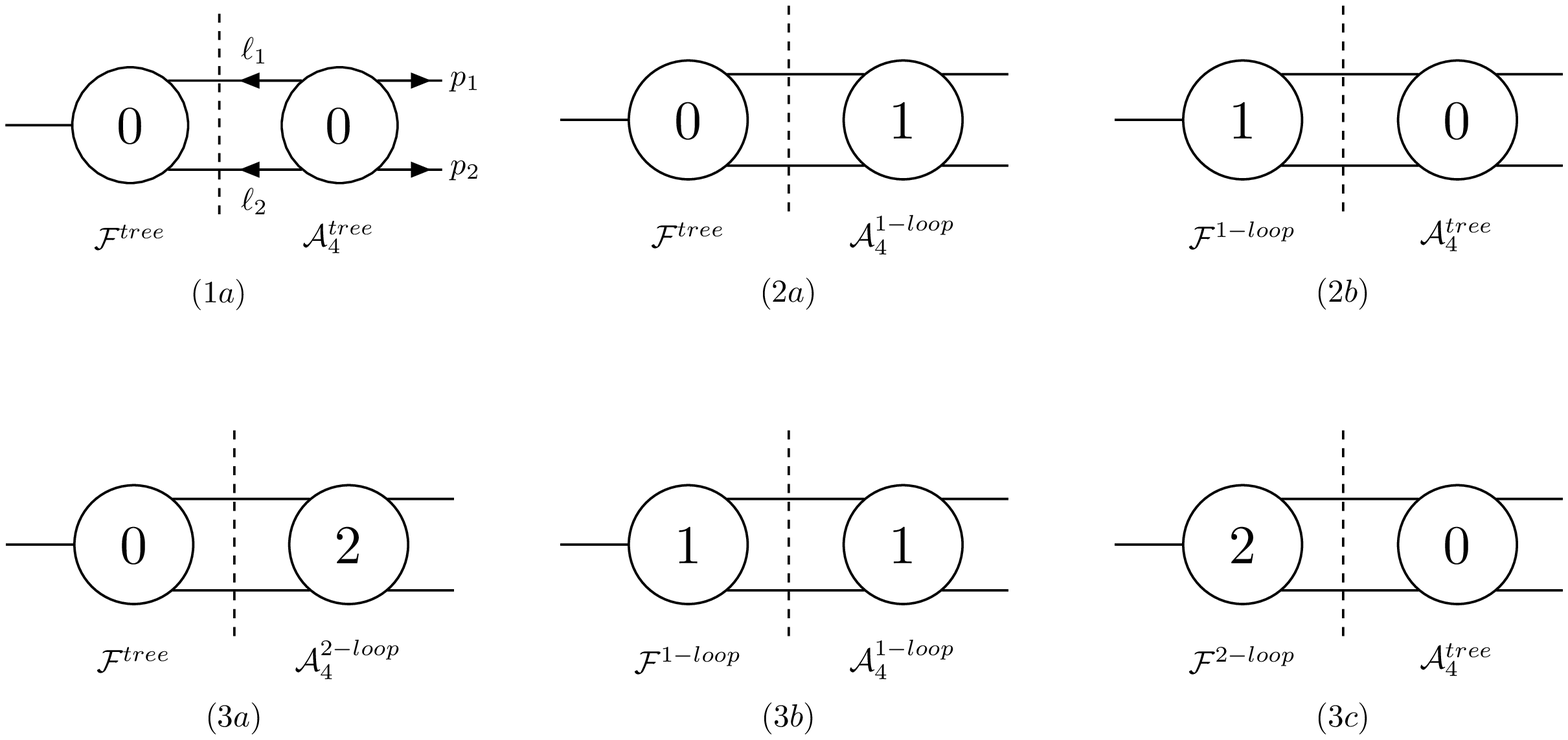}
\caption{Two-particle cuts of form factors up to three loops.}
\label{fig:2pcuts}}

As a simple warmup exercise, we rederive the one-loop result from unitarity cuts, see also ref. \cite{Brandhuber:2010ad}.
Let us compute the two-particle cut (1a) shown in Fig.~\ref{fig:2pcuts}.
It is given by
\be\label{eq:cut1loop}
{\cal F}_{S}^{1-loop} \Bigr|_{{\rm cut (1a)}} =  \int \sum_{P_1 , P_2 }  \frac{d^{D}k}{(2\pi)^{D}} \, \frac{i}{\ell_2^2} \, {\cal F}_{S}^{tree}(-\ell_1 , -\ell_2 ) \,  \frac{i}{\ell_1^2} \, \mathcal{A}_{4}^{tree}(\ell_2 , \ell_1, p_1 , p_2 )  \Bigr|_{\ell_1^2 = \ell_2^2=0} \,,
\ee
where $\ell_1$ and $\ell_2$ are the momenta of the cut legs, and the sum runs over all possible particles across the cut.
We may use the on-shell condition $\ell_1^2 = \ell_2^2  =0$ in the integrand (but not on the cut propagators), since any terms proportional to such numerator factors would vanish in the cut.
The four-particle tree ampliutde $\mathcal{A}_{4}^{tree}(\ell_2 , \ell_1, p_1 , p_2 )$ is given in
appendix~\ref{app:4pt}.  We use the convention that all momenta are defined as outgoing.

When computing the cut of a form factor (as opposed to a colour-ordered amplitude), 
one has to be careful about the overall normalisation, since the possible exchange of 
external legs $p_1$ and $p_2$ leads to a factor of $2$ in the cuts. When comparing cuts
of the form factor to cuts of integrals, this factor cancels out. In the following we  
count such contributions only once.

The two-particle cuts are particularly simple to evaluate. With our choice of external states, only
scalars can appear as intermediate particles, and we therefore do not need to use the spinor helicity
formalism. 
The tree-level form factor is simply given by
\be
{\cal F}_{S}^{tree}(-\ell_1 , -\ell_2 ) = \Tr(T^{a} T^{b}) \,.
\ee
The necessary four-particle amplitudes are given in appendix~\ref{app:4pt}.
The colour algebra across the cut is carried out using the $SU(N)$ Fierz identities, see eqs. (\ref{Fierz1}) and (\ref{Fierz2}).
It is easy to see that (\ref{eq:cut1loop}) becomes
\bea\label{eq:cut1loopb}
{\cal F}_{S}^{1-loop} \Bigr|_{{\rm cut\,(1a)}} &=& 
g^2 \mu^{2 \eps}
\, N  \, s_{12} \, \Tr(T^{a} T^{b} ) \,  
\int  \frac{d^{D}k}{(2\pi)^{D}} 
\, \frac{i}{\ell_2^2}  \,  \frac{i}{\ell_1^2} \, \left( \frac{-i}{(p_{1}+\ell_{1})^2} +\frac{-i}{(p_{2}+\ell_{2})^2}  \right)  \Bigr|_{\ell_1^2 = \ell_2^2=0} \,, \nonumber \\
&=& -2\,  g^2 \mu^{2 \eps}\, N \,  s_{12} \, \Tr(T^{a} T^{b} )\, 
\int  \frac{d^{D}k}{i (2\pi)^{D}} \frac{1}{k^2 (k+p_{1})^2 (k-p_{2})^2} \,
 \Bigr|_{{\rm cut\,(1a)}} \,,
\eea
where $s_{ij} := (p_{i}+p_{j})^2$, and where we have identified the cut of the one-loop form factor with
the cut of the one-loop triangle integral $D_{1}$, see Fig.~\ref{fig:diagsDE},
\be
D_1 = \int \frac{d^{D}k}{i (2 \pi)^D } \frac{1}{k^2 (k+p_{1})^2 (k-p_{2})^2}\,.
\ee
We can now argue that this result is exact, i.e.\ that we can remove the ``cut (1a)'' in Eq.~(\ref{eq:cut1loopb}).
In order to do that, we have to make sure that no terms with vanishing cuts are missed.
Such terms having no cuts in four dimensions can be detected in $D$ dimensions.
The two-particle cut calculation we just presented would have gone through unchanged in $D$ dimensions, since all required amplitudes were those of scalars,
and no spinor helicity identities intrinsic to four dimensions were used. 
A similar argument was given in ref. \cite{hep-ph/9702424}.
Therefore we conclude that in $D$ dimensions,
\be
{F}_{S}^{1-loop} = g^2  N \mu^{2 \eps} (-q^2) 2 D_1  \,.
\ee

\subsection{Two-loop form factor from unitarity cuts}

{}
\FIGURE[t]{
\includegraphics[width=0.95\textwidth]{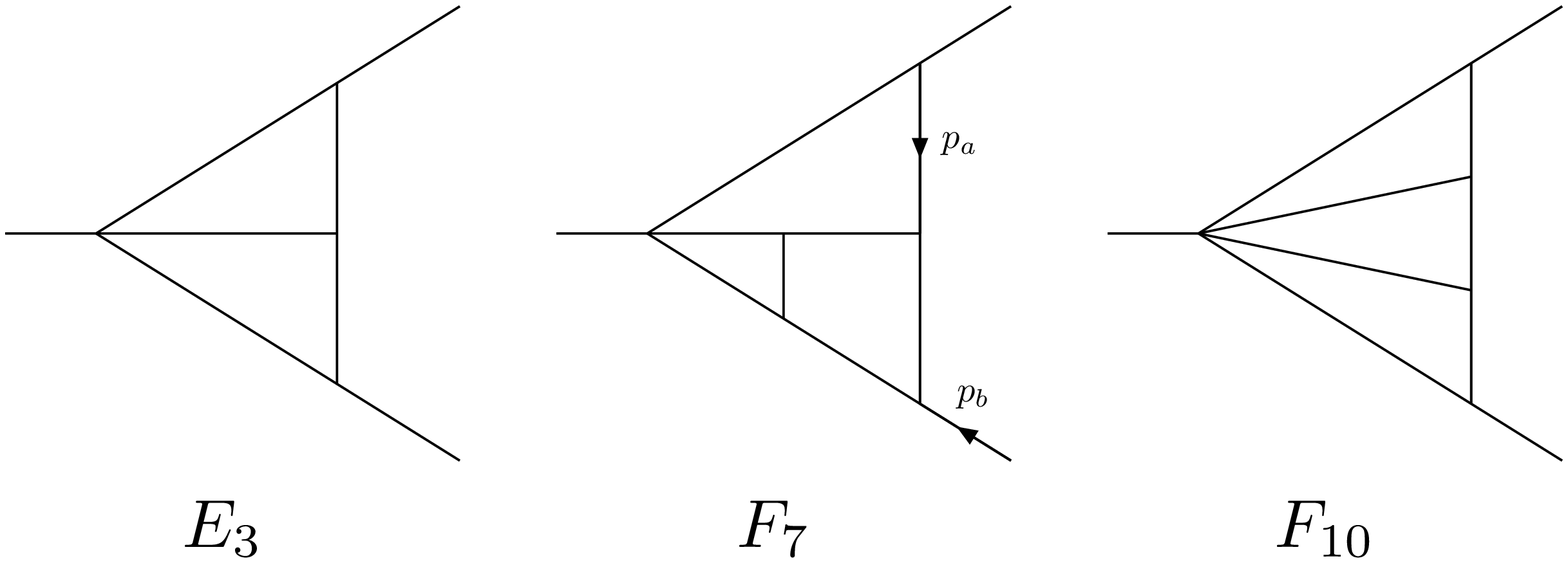}
\caption{Diagrams that do not contribute to the form factor
at two ($E_3$) and three loops ($F_{7}$ and $F_{10}$), respectively.
They have worse UV properties compared to the
integrals that do appear in the form factor.
The labels $p_a$ and $p_b$ on $F_{7}$ indicate an irreducible scalar product
$(p_a+p_b)^2$ in its numerator. The other two diagrams have unit numerator.}
\label{fig:34pcuts}}

We recall that at two loops, the result for the form factor is given by \cite{vanNeerven:1985ja},
\be\label{eq:vanNeerven}
{F}_{S}^{2-loop} = g^4 N^2 
\mu^{4 \eps} (-q^2)^2 \big\lbrack 4 E_1 + E_2 \big\rbrack  \,,
\ee
where the planar and non-planar ladder diagrams $E_1$ and $E_2$ are shown in Fig.~\ref{fig:diagsDE}.

Let us now understand this result from unitarity cuts.
The unitarity cut (2b) of Fig.~\ref{fig:2pcuts} detects the presence of the planar integral $E_1$ only.
The calculation is identical to that of the one-loop case, with the exception that the one-loop
form factor as opposed to the tree-level form factor is inserted on the l.h.s. of the cut.

The unitarity cut (2a) of Fig.~\ref{fig:2pcuts} reveals a new feature, that was already mentioned
in the introduction. On the r.h.s. of the cut we now insert the full one-loop four-point amplitude
${\cal A}_{4}^{1-loop}$, given explicitly in Eq.~(\ref{app:4pt1loop}), which in addition 
to single trace terms also contains double trace terms. The latter would ordinarily be subleading 
in the expansion of powers of $N$, e.g.\ when computing a four-point amplitude at leading
colour using unitarity cuts. Here, however the colour algebra gives rise to another
factor of $N$ for those terms, so that they can contribute to the form factor at the same order
as the single trace terms. This explains why the non-planar integral $E_2$ can appear in
the form factor.

In principle, new terms could appear in the three-particle cut, but this is not the case.
For example, the diagram $E_3$ shown in Fig.~\ref{fig:34pcuts} has no two-particle cuts.
The absence of this diagram can be understood by the fact that it has worse UV 
properties compared to $E_{1}$ and $E_{2}$, as we discuss in section \ref{sec:uv}.
For the same reason, diagrams $F_7$ and $F_{10}$ from
Fig.~\ref{fig:34pcuts}, the latter of which at has no three-particle
cuts, will not contribute to the form factor at three loops,  as we will see below.

We have also evaluated the three-particle and a generalised cut,
with the result being in perfect agreement with Eq.~(\ref{eq:vanNeerven}). 
We found it useful to employ a manifestly supersymmetric version of
the unitarity method \cite{Drummond:2008bq}. The necessary 
tree-level amplitudes for the local operator of Eq.~(\ref{eq:composite})
inserted into three on-shell states were computed in refs. 
\cite{Brandhuber:2010ad,Brandhuber:2011tv}.
The analytical calculation is straightforward to perform.
We refrain from presenting the details since it would require 
introducing spinor helicity and superspace.
We refer the interested reader to refs. \cite{Drummond:2008bq,Carrasco:2011hw} for
related instructive examples.

\subsection{Three-loop form factor from unitarity cuts}

We again begin by studying two-particle cuts, which are shown in 
the second line of Fig.~\ref{fig:2pcuts}.
Again, all results for the form factors and four-point amplitudes appearing in
the unitarity cuts are explicitly known, with the result for the four-point amplitudes
summarized in appendix \ref{app:4pt}.

When evaluating the cuts, one has a certain freedom in rewriting the answer to
a given cut due to the on-shell conditions. Of course, eventually such ambiguities
are fixed by the requirement that the answer must satisfy all cuts. 
In order to find such an expression that manifestly satisfies all cuts it is very
useful to have an idea about the kind of integrals that should appear in the 
answer. We expect that the form factor can be expressed in terms of
the integrals that have UT that were discussed in section \ref{sec:rpi}.
This turns out to be a very useful guiding principle.

The calculation is completely analogous to that at one and two loops.
Let us start with the simplest cut (3c) from Fig.~\ref{fig:2pcuts}. It is given by
\be\label{eq:cut3loop3c}
{\cal F}_{S}^{3-loop} \Bigr|_{{\rm cut (3c)}} =  \int \sum_{P_1 , P_2 }  \frac{d^{D}k}{(2\pi)^{D}} \, \frac{i}{\ell_2^2} \, {\cal F}_{S}^{2-loop}(-\ell_1 , -\ell_2 ) \,  \frac{i}{\ell_1^2} \, \mathcal{A}_{4}^{tree}(\ell_2 , \ell_1, p_1 , p_2 )  \Bigr|_{\ell_1^2 = \ell_2^2=0} \,,
\ee
The evaluation of the cut is exactly as that considered at one loop, with the difference that
we now insert the two-loop expression for the form factor into the cut, as opposed to the tree-level one.
One immediately finds
\be
{\cal F}_{S}^{3-loop} \Bigr|_{{\rm cut (3c)}} = g^6\,\mu^{6\epsilon}\,N^3 \, (-q^2)^3 \big\lbrack 8 \, F_{1} + 2\,  F_{3}^{*} \big\rbrack \Bigr|_{{\rm cut (3c)}}  \,,
\ee
where $F_{1}$ is the three-loop ladder integral shown in Fig.~\ref{fig:diagsF},
and $F_{3}^{*}$ is related to $F_{3}$ in the same figure via the identity (\ref{eq:RPIF3}).
In fact, we know from section \ref{sec:rpi} that $F_{3}^{*}$ does not have uniform transcendentality.
Since we do expect the final result to have this property, use Eq.~(\ref{eq:RPIF3}) to
eliminate $F_{3}^{*}$. When doing so, we note that the contribution of $F_{a1}$ in
that equation drops out on the cut (3c), and we have
\be\label{result-cut3c}
{ F}_{S}^{3-loop} \Bigr|_{{\rm cut (3c)}} = g^6\,\mu^{6\epsilon}\,N^3 \, (-q^2)^2 \big\lbrack 8 \,(-q^2) \, F_{1} + 4  \, F_{3} - 4 \, F_{8}  \big\rbrack \Bigr|_{{\rm cut (3c)}} \,,
\ee
i.e.\ we have succeeded in writing the two-particle cut (3c) in terms of integrals having UT only.

Similarly, one can show that the two-particle cut (3b) of Fig.~\ref{fig:2pcuts} can be written as
\be\label{result-cut3b}
{ F}_{S}^{3-loop} \Bigr|_{{\rm cut (3b)}} = g^6\,\mu^{6\epsilon}\,N^3 \, (-q^2)^2 \big\lbrack 8 \,(-q^2) \, F_{1} + 4  \,  F_{4}   \big\rbrack \Bigr|_{{\rm cut (3b)}} \,.
\ee
This confirms the coefficient of $F_{1}$, and introduces a new integral $F_{4}$, invisible to cut (3c).

Finally, the most interesting two-particle cut is (3a), as it uses the double trace terms
present in $\mathcal{A}_{4}^{2-loop}$, see appendix \ref{app:4pt}. Using the identities
derived in section \ref{sec:rpi}, we find
\be\label{result-cut3a}
{ F}_{S}^{3-loop} \Bigr|_{{\rm cut (3a)}} = 
g^6\,\mu^{6\epsilon}\,N^3 \, (-q^2)^2 \big\lbrack 
8 \,(-q^2)\,  F_{1} - 2 \,  F_{2} + 4\,  F_{3} + 4 \,  F_{4} -4 \,  F_{5} - 4\,  F_{6}  \big\rbrack \Bigr|_{{\rm cut (3a)}} \,.
\ee

Comparing equations (\ref{result-cut3c}),(\ref{result-cut3b}), and (\ref{result-cut3a}) with each 
other, we see that they are manifestly consistent with each other, 
which suggests that we are indeed working with an appropriate integral basis to describe this problem.
We find that the following expression is in agreement with all two-particle cuts,
\be\label{result-2pcuts}
{ F}_{S}^{3-loop} \Bigr|_{{\rm 2-part.\; cut}} = 
g^6\,\mu^{6\epsilon}\,N^3 \, (-q^2)^2 \big\lbrack 
8\, (-q^2) \,  F_{1} - 2\,  F_{2} + 4\,  F_{3} + 4  \, F_{4} -4\,  F_{5} - 4 \, F_{6} - 4\,  F_{8} \big\rbrack \Bigr|_{{\rm 2-part.\; cut}} \,.
\ee
It is quite remarkable that to three loops the coefficients of all
integrals are small integer numbers.

We could proceed by evaluating three- and four-particle cuts, but we find it
technically simpler to study generalised cuts.
To begin with, we perform a cross-check on the two-particle cut
calculation above by evaluating maximal cuts where nine propagators are
cut. We find perfect agreement between the two calculations.
Next, we release one cut constraint to detect integrals having only 
eight propagators. There are several ways in which this can be done.
For example, cutting all eight propagators present in integral $F_{9}$
detects this integral, as well as integrals $F_{5}$ and $F_{6}$.
Another eight-propagator cut detects integrals $F_{2}, F_{5}, F_{6}$ and $F_{7}$.
The latter integral (see Fig.~\ref{fig:34pcuts}) turns out to have
coefficient zero, i.e.\ it does not appear.

We again find perfect agreement with the contributions already known from the
two-particle cuts, and find further contributions not having any
two-particle cuts, like $F_9$.
The following expression satisfies all cuts that we have evaluated,
\bea\label{eq:3lfirst}
{F}_S^{3-loop} &=&   g^6 \, \mu^{6\eps}  \, N^3 \, (-q^2)^2 \,
\big\lbrack 8 \, (-q^2) \, F_1 - 2 \, F_2 + 4 \, F_3 + 4 \, F_4 - 4 \,
F_5 - 4 \, F_6 - 4 \, F_8 + 2 \, F_9 \big\rbrack \,. \nonumber \\ 
\eea
We will now argue that Eq.~(\ref{eq:3lfirst}) is the complete result for the three-loop form factor.
In fact, potential corrections to equation (\ref{eq:3lfirst}) can come only from seven-propagator 
integrals that have vanishing two-particle cuts. An example of such an integral is
$F_{10}$ shown in Fig.~\ref{fig:34pcuts}.
As we will see in section \ref{sec:uv}, the appearance of such integrals is highly 
unlikely due to their bad UV behaviour, violating a bound based on supersymmetry power counting.

Moreover, in section \ref{sec:logff}, we will perform an even more stringent 
check on Eq.~(\ref{eq:3lfirst}) by verifying the correct exponentiation of infrared divergences.
In particular, this means that any potentially missing terms 
in equation (\ref{eq:3lfirst}) would have to be IR and UV finite,
and vanish in all unitarity cuts that we considered.


\section{Final result for the form factor at three loops}\label{sec:ff3l}

{}
\FIGURE[t]{
\includegraphics[width=0.99\textwidth]{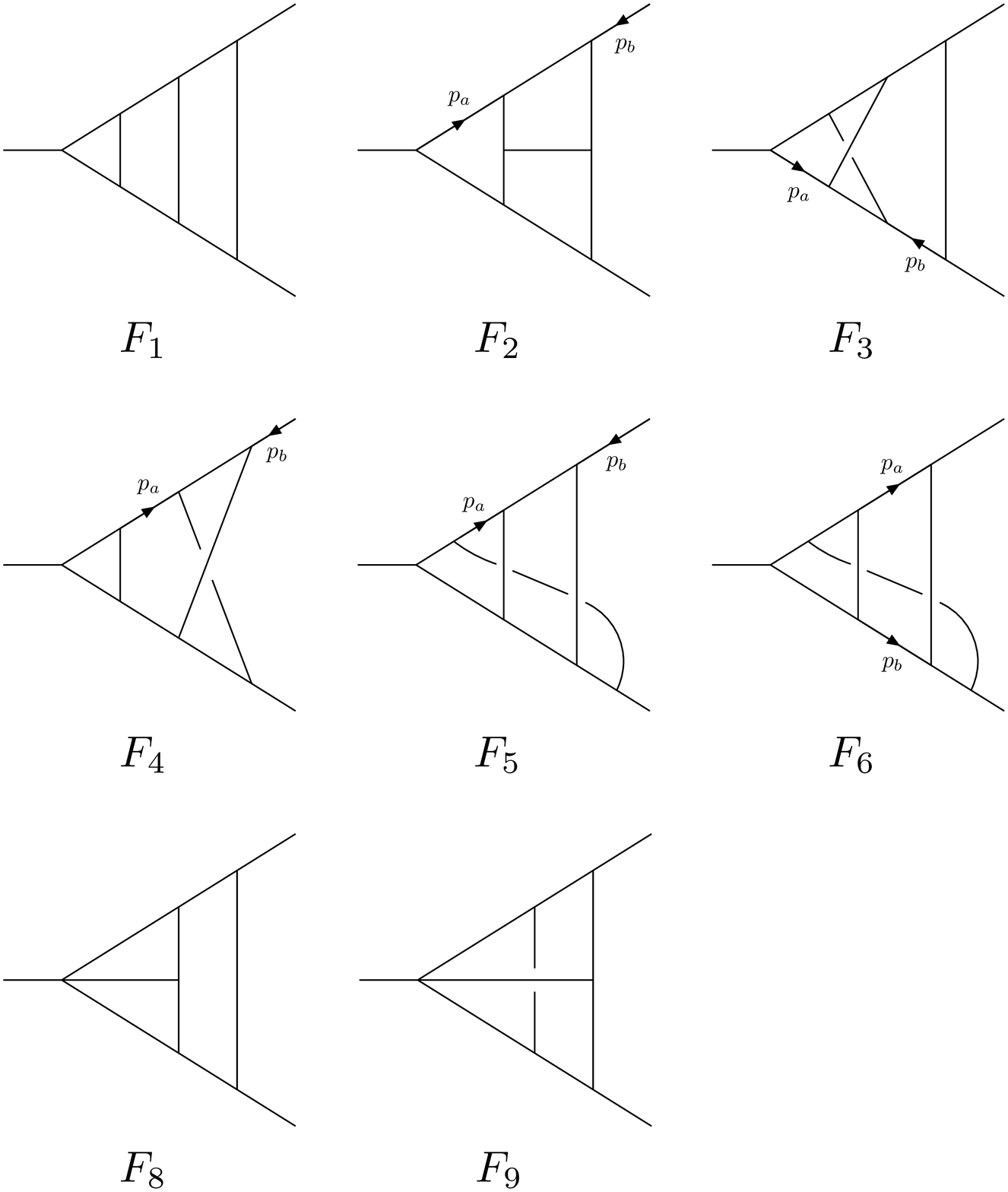}
\caption{Diagrams of which the three-loop form factor
${F}_S^{(3)}$ in $\mathcal{N}=4$ SYM is built.
All internal lines are massless. The incoming momentum is
$q=p_1+p_2$, outgoing lines are massless and on-shell,
i.e.\ $p_1^2 = p_2^2 = 0$. Diagrams with labels $p_a$ and
$p_b$ on arrow lines have an irreducible scalar product
$(p_a+p_b)^2$ in their numerator (diagrams that lack these
labels have unit numerator). All diagrams displayed exhibit
uniform transcendentality (UT) in their Laurent expansion
in $\eps = (4-D)/2$.}
\label{fig:diagsF}}

In the previous section we 
obtained the extension of Eq.~(\ref{eq:ffexpansion}) to three loops,
{\allowdisplaybreaks
\bea
{F}_S &=& 1+ g^2 \, N \, \mu^{2\eps} \cdot (-q^2)\cdot 2\, D_1
+ g^4 \, N^2 \, \mu^{4\eps} \cdot (-q^2)^2 \cdot
\left[4 \, E_1 + E_2\right]  \nnb \\
&& + \, g^6 \, N^3 \, \mu^{6\eps} \cdot (-q^2)^2 \cdot
\left[8 \, (-q^2) \, F_1 - 2 \, F_2 + 4 \, F_3 + 4 \, F_4 - 4 \, F_5 - 4 \, F_6 - 4 \, F_8 + 2 \, F_9\right] \nnb \\
&& + \, {\cal{O}}(g^8) \nnb \\
&=& 1+ a \, x^{\eps} \, R_{\eps} \cdot 2 \, D_1^{{\rm exp}}
+ a^2 \, x^{2\eps} \, R_{\eps}^2
\cdot \left[4 \, E_1^{{\rm exp}} + E_2^{{\rm exp}}\right] \nnb \\
&& + a^3 \, x^{3\eps} \, R_{\eps}^3 \cdot
\left[8 \, F_1^{{\rm exp}} - 2 \, F_2^{{\rm exp}} + 4 \, F_3^{{\rm exp}}
+ 4 \, F_4^{{\rm exp}} - 4 \, F_5^{{\rm exp}} - 4 \, F_6^{{\rm exp}} - 4 \, F_8^{{\rm exp}} + 2 \, F_9^{{\rm exp}}\right] \nnb \\
&& +{\cal{O}}(a^4)\; . \label{eq:3lut}
\eea
}
The expressions for $F_i$, and $F_i^{{\rm
exp}}$ are again given in appendix~\ref{app:integrals}. All diagrams
are displayed in Fig.~\ref{fig:diagsF}. This yields
{\allowdisplaybreaks
\bea
{F}_S^{(3)} &=& R_{\eps}^3
\cdot \left[8 \, F_1^{{\rm exp}} - 2 \, F_2^{{\rm exp}} + 4 \, F_3^{{\rm exp}}
+ 4 \, F_4^{{\rm exp}} - 4 \, F_5^{{\rm exp}} - 4 \, F_6^{{\rm exp}} - 4 \, F_8^{{\rm exp}} + 2 \, F_9^{{\rm exp}}\right] \nnb \\
&=& -\frac{1}{6 \eps^6}+\frac{11 \zeta_3}{12 \eps^3}+\frac{247 \pi ^4}{25920 \eps^2}
+\frac{1}{\eps}\left(-\frac{85 \pi ^2 \zeta_3}{432}-\frac{439 \zeta_5}{60}\right) \nnb \\
&& -\frac{883 \zeta_3^2}{36}-\frac{22523 \pi ^6}{466560}
+\eps \left(-\frac{47803 \pi ^4 \zeta_3}{51840}+\frac{2449 \pi ^2 \zeta_5}{432}-\frac{385579 \zeta_7}{1008}\right) \nnb \\
&& + \eps^2 \left(\frac{1549}{45} \zeta_{5,3}-\frac{22499 \zeta_3 \zeta_5}{30}+\frac{496 \pi ^2 \zeta_3^2}{27}
-\frac{1183759981 \pi ^8}{7838208000}\right) + {\cal{O}}(\eps^3) \; . \label{eq:3lff}
\eea
}

We can make a very interesting observation here.
For anomalous dimensions of twist two operators, there is a heuristic leading transcendentality 
principle~\cite{hep-ph/0112346,hep-th/0404092,hep-th/0611204}, which relates
the $\cN=4$ SYM result to the  leading transcendental part of the QCD result.
We can investigate whether a similar property holds for the form factor.

For the comparison, we specify the QCD quark and gluon form
factor to a supersymmetric Yang-Mills theory containing a bosonic and
fermionic degree of freedom in the same colour representation, which is
achieved by setting $C_A = C_F = 2 \, T_F$ and $n_f=1$ in the QCD
result~\cite{BCSSS}. It turns out that with this adjustment the leading
transcendentality pieces of the quark and gluon form factor become
equal at one, two, and three loops in all coefficients up to
transcendental weight eight, i.e.\ ${\cal O}(\eps^6)$, ${\cal
O}(\eps^4)$, and ${\cal O}(\eps^2)$ at one, two, and three loops,
respectively. Moreover, the leading transcendentality pieces of the
quark and gluon form factor coincide -- up to a factor of $2^L$ ($L$ is
the number of loops) which is due to normalisation -- with the
coefficients of the scalar form factor in $\cN=4$~SYM computed in the
present work. This again holds true at one, two, and three loops and for
all coefficients up to weight eight, and serves as an important check
of our result.

The question arises if the leading transcendentality
principle~\cite{hep-ph/0112346,hep-th/0404092,hep-th/0611204} 
between QCD and $\cN=4$
SYM carries over to more general quantites like scattering amplitudes,
or if it is a special feature of form factors since they have only two
external partons. 

In fact, there are counterexamples in the case of scattering
amplitudes~\cite{Lance}. 
For instance, the $\cN=1$ supersymmetric one-loop four-point amplitudes \cite{Kunszt:1993sd}
have a leading transcendentality piece which is not of the $\cN=4$ SYM form, because it has
$1/u$ power-law factors. 
This makes the property we have found for the form factor even more surprising.


\section{Logarithm of the form factor}\label{sec:logff}

The logarithm of the form factor is given by

{\allowdisplaybreaks
\bea
\dps \ln\left({F}_S\right) &=& 
\ln\left(1+ a \, x^{\eps} \, {F}_S^{(1)}
+ a^2 \, x^{2\eps} \, {F}_S^{(2)}
+ a^3 \, x^{3\eps} \, {F}_S^{(3)}+{\cal{O}}(a^4)\right) \nnb\\
&=& a \, x^{\eps} \, {F}_S^{(1)}
+ a^2 \, x^{2\eps} \left[{F}_S^{(2)}-\frac{1}{2} \left({F}_S^{(1)}\right)^2\right]
+ a^3 \, x^{3\eps} \left[{F}_S^{(3)}-{F}_S^{(1)}{F}_S^{(2)}
+\frac{1}{3}\left({F}_S^{(1)}\right)^3\right] \nnb\\
&&+{\cal{O}}(a^4) \; ,
\eea
}
where
{\allowdisplaybreaks
\bea
{F}_S^{(1)} &=& -\frac{1}{\eps^2}+\frac{\pi ^2}{12}+\frac{7\zeta_3}{3} \, \eps
+ \frac{47 \pi ^4}{1440} \, \eps^2 +\eps^3 \left(\frac{31
\zeta_5}{5}-\frac{7 \pi ^2\zeta_3}{36}\right) \nnb \\
&& +\eps^4 \left(\frac{949\pi ^6}{120960}-\frac{49\zeta_3^2}{18}\right)
+\eps^5\left(\frac{127\zeta_7}{7}-\frac{329\pi ^4 \zeta_3}{4320}-\frac{31 \pi ^2 \zeta_5}{60}\right)  \nnb \\
&& +\eps^6 \left(\frac{49 \pi ^2\zeta_3^2}{216}-\frac{217 \zeta_3 \zeta_5}{15}+\frac{18593 \pi^8}{9676800}\right)
+ {\cal{O}}(\eps^7) \; , \\
&& \nnb \\
{F}_S^{(2)}-\frac{1}{2} \left({F}_S^{(1)}\right)^2 &=&
\frac{\pi ^2}{24 \eps^2}+\frac{\zeta_3}{4 \eps} +\eps \left(\frac{39 \zeta_5}{4}-\frac{5 \pi ^2 \zeta_3}{72}\right)
+\eps^2 \left(\frac{235 \zeta_3^2}{12}+\frac{2623 \pi ^6}{60480}\right)\nnb \\
&&+\eps^3 \left(\frac{73 \pi ^4 \zeta_3}{96}-\frac{437 \pi ^2 \zeta_5}{120}+\frac{489 \zeta_7}{2}\right)\nnb \\
&&+\eps^4 \left(-66 \zeta_{5,3}+\frac{1119 \zeta_3 \zeta_5}{10}-\frac{1351 \pi ^2 \zeta_3^2}{216}
+\frac{127 \pi ^8}{1296}\right)+ {\cal{O}}(\eps^5) \; , \nnb \\
&& \\
{F}_S^{(3)}-{F}_S^{(1)}{F}_S^{(2)}
+\frac{1}{3}\left({F}_S^{(1)}\right)^3 &=&-\frac{11 \pi ^4}{1620 \eps^2}
+\frac{1}{\eps}\left(-\frac{5 \pi ^2 \zeta_3}{54}-\frac{2 \zeta_5}{3}\right)
-\frac{13 \zeta_3^2}{9}-\frac{193 \pi ^6}{25515} \nnb \\
&&+\eps \left(-\frac{107 \pi ^4 \zeta_3}{1620}+\frac{187 \pi ^2 \zeta_5}{108}-\frac{21181 \zeta_7}{144}\right) \nnb \\
&&+\eps^2 \left(-\frac{1421}{45} \zeta_{5,3}-\frac{1922 \zeta_3 \zeta_5}{3}+\frac{1057 \pi ^2 \zeta_3^2}{108}-\frac{994807 \pi ^8}{17496000}\right)
\nnb \\
&&+ {\cal{O}}(\eps^3) \; .
\eea
}
The poles of the logarithm of the form factor have the generic
structure~\cite{Magnea:1990zb}
\be
\ln\left({F}_S\right) =
\sum\limits_{L=1}^{\infty} a^L \, x^{L\eps} \,
\left[ 
-\frac{\gamma^{(L)}}{ 4 (L \eps)^2} 
-\frac{\cG_0^{(L)}} {2 L \eps} 
\right] + {\cal{O}}(\eps^0) \; ,
\ee
with the $L$-loop cusp $\gamma^{(L)}$    
and collinear $\cG_0^{(L)} $ 
anomalous dimensions~\cite{Korchemskaya:1992je}
given by
\bea
\label{eq:gammacusp}
\gamma(a) 
&=& 
\sum\limits_{L=1}^\infty a^L  
\gamma^{(L)}
= 
4a - 4\zeta_2 a^2 + 22 \zeta_4 a^3 + {\cal{O}}(a^4) \; ,
\\
\label{eq:gammaG}
\cG_0(a)
&=& 
\sum\limits_{L=1}^\infty  a^L  \cG_0^{(L)} =
    -  \zeta_3 a^2 + \left(4 \zeta_5 + \frac{10}{3} \zeta_2 \zeta_3
    \right)  a^3 + {\cal{O}}(a^4) \; .
\eea


We observe that the vanishing of the ${\cal O}(\epsilon^0)$-term in the 
logarithm of the two-loop form factor~\cite{vanNeerven:1985ja} appears to 
be a coincidence, which does not reproduce at three loops. The finite 
part of the ${\cal N}=4$ form factor does therefore not exponentiate, 
as could have been conjectured from the two-loop result.

\section{Ultraviolet divergences in higher dimensions}\label{sec:uv}

Scattering amplitudes and form factors in $\cN=4$ super Yang-Mills are ultraviolet (UV) finite
in four dimensions.
It is interesting to ask in what dimension, called critical dimension $D_{c}$, 
they first develop UV divergences. 
This question is of theoretical interest in the context of the discussion of possible
finiteness of $\cN=8$ supergravity, see e.g.\ \cite{Bern:2009kd} and references therein. More practically,
bounds on the critical UV dimension at a given loop order
can also be a useful cross-check of computations, or constrain the types of loop
integrals that can appear.

There is a  bound on the critical dimension based on power counting for supergraphs and the background field 
method.
The one-loop case is special due to some technical issue with ghosts, 
but there is a bound for $L>1$ loops \cite{Grisaru:1982zh,Marcus:1984ei},
\be \label{susyboundUV}
D_{c}(L) \ge 4 + \frac{2 (\cN-1)}{L} \,, \qquad L>1\,,
\ee
such that for $D<D_{c}$ the theory is UV finite.
The bound (\ref{susyboundUV}) depends on the number $\cN$ of supersymmetries that can be realized off-shell.
The maximal amount of supersymmetry can be realised using an $\cN=3$ harmonic superspace 
action for $\cN=4$ super Yang-Mills~\cite{570842}.
Taking thus $\cN=3$ in (\ref{susyboundUV}) we have
\be  \label{susyboundUV2}
D_{c}(L) \ge 4 + \frac{4}{L} \,,\qquad L>1\,.
\ee
 
Equation (\ref{susyboundUV}) is a lower bound for $D_{c}$, and in some cases it can be too conservative.
For example, in the case of scattering amplitudes, studying and excluding potential counterterms 
bounds on the critical dimension can sometimes be improved, see the reviews \cite{Bossard:2009sy,Howe:1988qz}.
Investigations of UV properties of four-particle scattering amplitudes have shown that
their ultraviolet behaviour is better than expected \cite{Bern:2010tq}. Their critical dimension at two and three loops
was shown to be $7$ and $6$, respectively, suggesting the improved bound $D_{c}(L) \ge 4 + {6}/{L}$.
The one-loop case is exceptional, but for completeness we note that $D_{c}(L=1)=8$ for the four-particle
scattering amplitude.

We can now study the UV properties for
$D>4$ of the form factor that we have computed. There is no statement
from Eq.~(\ref{susyboundUV2}) for the one-loop case, but one can
easily see that $D_{c}(L=1) = 6$. For the two-loop form factor, the
bound (\ref{susyboundUV2}) is actually saturated since the two-loop
form factor develops its first ultraviolet divergence at $D_{c}(L=2) =
6$. Moreover, it turns out that in $D=6-2\eps$ dimensions the leading
$1/\eps^2$ UV-pole is given by the leading UV-pole of the two-loop
planar ladder diagram $E_1$, and that $E_2$ has only a simple $1/\eps$ pole.

At three loops,  Eq.~(\ref{susyboundUV2}) becomes $D_{c} \ge 16/3$.
First of all, we see by power counting that diagrams $F_{7}$ and
$F_{10}$ (see Fig.~\ref{fig:34pcuts}) both have a UV divergence in $D= 14/3$ dimensions,
which would violate the supersymmetry bound (\ref{susyboundUV2}). 
This comes close to explaining why their coefficients are zero, and why
other integrals having seven or fewer propagators do not appear. A small
caveat is that it may not always be possible to write the answer in a form
such that the UV properties are manifest: one could have
a linear combination of integrals that individually have worse UV properties 
than expected, but with appropriate UV behaviour of the linear combination.
However, as we will see presently, we can make the UV properties of the three-loop
form factor completely manifest.

At two loops we found that the bound from superspace counting was saturated.
We can ask whether the same happens at three loops, i.e.\ 
do we have ${D_{c}(L=3) = 16/3}$?
It turns out that the three-loop form factor is better
behaved in the UV than suggested by this equation.
It is finite in $D = 16/3$ and only develops a UV divergence
at $D_{c}(L=3) =6$. In order to see this, we take the three-loop
expression~(\ref{eq:3lfirst}) and trade $F_3$, $F_4$ and $F_5$ for the
non-UT integrals $F_3^*$, $F_4^*$ and $F_5^*$ by means of
Eqs.~(\ref{eq:RPIF3}), (\ref{eq:RPIF4}), and  (\ref{eq:RPIF5}),
respectively, which leads to
\be\label{eq:FFUV}
{F}_S^{3-loop} \propto
(-q^2) \, \left[ 8 \, F_1 + 2 \, F_3^* + 2 \, F_4^*\right] - 2 \, F_2 + 4 \, F_5^* - 2 \, F_9 \,.
\ee
Counting numerators as propagators with negative powers, we see that the
three integrals in the bracket have nine propagators each, whereas the
last three integrals have only eight propagators. Since there are no
sub-divergences in $D=16/3$ we can calculate the leading UV pole by
simply giving all propagators (and also all numerators\footnote{Whether
or not we give a mass to the numerators changes the expressions
only by integrals with nine propagators each. The latter are finite in
$D=16/3$ by na\"ive power counting.}) a
common mass $m$ and by setting the external momenta $p_1=p_2=0$. Then
the first three integrals are finite by na\"ive power counting, and the
last three integrals become equal, and cancel due to their pre-factors.
This renders the three-loop form factor finite in $D=16/3$ dimensions.
One can see the UV finiteness of
the $\cN=4$ SYM form factor in $D=16/3$ also in another, more elegant
way. We start again from Eq.~(\ref{eq:FFUV}), and add zero in the
disguise of
\be
+ \, 2 \, F_7^* \, - \, 2 \, F_7^* \; ,
\ee
where $F_7^*$ is $F_7$ (see Fig.~\ref{fig:34pcuts}) with unit numerator.
This choice is particularly convenient since $F_7^*$ is a subtopology of
both, $F_2$ and $F_5^*$. It is obtained from $F_2$ by shrinking the line
labelled $p_a$ in Fig.~\ref{fig:diagsF}. Alternatively, $F_7^*$ is obtained from
$F_5^*$ by shrinking line number 7 in Fig.~\ref{fig:diagsRPI3}. In both
cases one subsequently has to set the respective numerator to unity.
Hence we can rewrite~(\ref{eq:FFUV}) as
\be\label{eq:FFUVmanifest}
{F}_S^{3-loop} \propto
(-q^2) \, \left[ 8 \, F_1 + 2 \, F_3^* + 2 \, F_4^*\right]
- 2 \left( F_2 - F_7^*\right) + 2 \left( 2 \, F_5^* - F_7^* - F_9 \right)\,.
\ee
If we adopt for $F_2$ the parametrisation
\be\label{eq:linesF2}
\left\{ k_1  \, , \, k_1 + p_1 \, , \, k_2 \, ,
 \, k_2 + p_2 \, , \, k_3 - p_2 \, , \, k_3 + p_1 \, ,
 \, k_1 +  k_2 \, , \, k_1 - k_3 \, , \, k_2 + k_3 \, \right\} \; , 
\ee
and write $\left(F_2 - F_7^*\right)$ on a common denominator, the
numerator of the latter expression reads
\be
k_3^2 - \left(k_3 - p_2\right)^2
\ee
and hence vanishes in the aforementioned UV limit. In complete analogy,
we take the parametrisation~(\ref{eq:linesF5star}) for $F_5^*$ and write
$\left( 2 \, F_5^* - F_7^* - \, F_9 \right)$ on a common denominator,
whose numerator becomes
\be
\left[\left(k_2-p_1\right)^2 - k_2^2 \right] + \left[\left(k_2-p_1\right)^2 - \left(k_2-p_1-p_2\right)^2 \right] \; ,
\ee
which clearly also vanishes upon taking the UV limit. Hence
Eqs.~(\ref{eq:FFUV}) and~(\ref{eq:FFUVmanifest}) make the UV properties of the form factor
manifest.
This is very similar to how the UV properties of four-particle amplitudes 
can be made manifest, see e.g.  ref. \cite{Bern:2010tq}.

It is now interesting to investigate the UV properties of the form
factor in $D=6-2\eps$ dimensions. Since the vanishing of $\left(F_2 -
F_7^*\right)$ and $\left( 2 \, F_5^* - F_7^* - \, F_9 \right)$ should be
independent of the number of dimensions, we can simply look at the
expression
\be
8 \, F_1 + 2 \, F_3^* + 2 \, F_4^* \,,
\ee
and the corresponding integrals at one and two loops. Introducing a common
propagator mass and neglecting external momenta one finds
{\allowdisplaybreaks
\bea
2 D_1^{\rm UV} &\stackrel{D=6-2\eps}{=}& \ESGamma
\left[m^2\right]^{-\eps} \, \left\{-\frac{1}{\eps}-\frac{\pi^2}{6} \,
\eps - \frac{7\pi^4}{360} \, \eps^3 + {\cal O}(\eps^5)\right\} \; ,
\nnb\\
4 E_1^{\rm UV}  + E_2^{\rm UV} &\stackrel{D=6-2\eps}{=}& \ESGamma^2
\left[m^2\right]^{-2\eps} \, \left\{\frac{1}{2\eps^2}+\frac{1}{2\eps}
+\left[\frac{1}{2} +\frac{\pi^2}{6} -\frac{1}{5} \, a_\Phi\right] + {\cal O}(\eps)\right\} \; ,
\nnb\\
8 F_1^{\rm UV}  + 2 F_3^{* \, \rm UV} + 2 F_4^{* \, \rm UV}&\stackrel{D=6-2\eps}{=}& \ESGamma^3
\left[m^2\right]^{-3\eps} \,\left\{-\frac{1}{6\eps^3}-\frac{1}{2\eps^2}\right.\nnb \\
&& \qquad \quad \left.+\frac{1}{\eps}
\left[\frac{\zeta_3}{3} -\frac{\pi^2}{12} -\frac{13}{9}+\frac{1}{5} \, a_\Phi \right]+
{\cal O}(\eps^0)\right\} \; , \label{eq:UVmass}
\eea}
where $\ESGamma$ is defined in appendix~\ref{app:integrals}, and (see
e.g.\ ~\cite{Huber:2009se})
\bea
a_\Phi &=& \Phi(\textstyle-\frac{1}{3},2,\frac{1}{2}) \displaystyle +
\frac{\pi\ln(3)}{\sqrt{3}} \; , \\
\Phi\!\left(z,s,a\right) &=& \sum\limits_{k=0}^{\infty} \frac{z^k}{[(k+a)^2]^{s/2}} \; ,\\
\Phi(\textstyle-\frac{1}{3},2,\frac{1}{2}) &=& 4 \, \sqrt{3} \; {\rm
Im}\!\!\left[\PL{2}{\textstyle\frac{i}{\sqrt{3}}}\displaystyle\right] 
= -\frac{\pi\ln(3)}{\sqrt{3}}+\frac{10}{\sqrt{3}} \; {\rm Cl}_2\!\!\left(\frac{\pi}{3}\right) \; ,
\eea
and $\rm Cl_2$ is the Clausen function. Hence we find that up to three
loops the form factor at each loop-order has $D_{c}=6$. Moreover, it
turns out that for $D=6-2\eps$ the leading $1/\eps^L$ UV-pole is at each
loop order given by the leading UV-pole of the respective $L$-loop
planar ladder diagram. Since at $D=6-2\eps$ there might be issues due to
the presence of sub-divergences, we also computed the UV divergences
using a different regulator. After having taken the soft limit, we
re-insert some external momentum into the graph to serve as IR
regulator, instead of the mass (essentially, one nullifies one of the
$p_i$ and takes the other one off-shell). In this way one obtains massless
propagator type integrals which lead to the following result
{\allowdisplaybreaks
\bea
2 D_1^{\rm UV} &\stackrel{D=6-2\eps}{=}& \ESGamma
\left(-q^2\right)^{-\eps} \, \left\{-\frac{1}{\eps}-2 - 4 \, \eps 
+ (2\zeta_3-8) \, \eps^2+ {\cal O}(\eps^3)\right\} \; ,
\nnb\\
4 E_1^{\rm UV}  + E_2^{\rm UV} &\stackrel{D=6-2\eps}{=}& \ESGamma^2
\left(-q^2\right)^{-2\eps} \, \left\{\frac{1}{2\eps^2}+\frac{5}{2\eps}
+\left[\frac{53}{6} - \zeta_3\right] + {\cal O}(\eps) \right\} \; , \nnb\\
8 F_1^{\rm UV}  + 2 F_3^{* \, \rm UV} + 2 F_4^{* \, \rm UV}&\stackrel{D=6-2\eps}{=}& \ESGamma^3
\left(-q^2\right)^{-3\eps} \,\left\{-\frac{1}{6\eps^3}-\frac{3}{2\eps^2}
+\frac{1}{\eps}\left[\frac{4\zeta_3}{3} -\frac{79}{9}\right]+{\cal O}(\eps^0)\right\}
\; . \nnb\\ \label{eq:UVmom}
\eea}
As expected, the leading $\eps^{-L}$ divergence at $L$ loops is independent of the
regulator, while the subleading terms are not. However, when considering
$\log(F_{\cal S})$ in the UV limit there are only simple $1/\eps$ poles up to three loops. 
Moreover, these poles are
identical in both regularisation schemes~(\ref{eq:UVmass})
and~(\ref{eq:UVmom}), and read
{\allowdisplaybreaks
\begin{align}
\ln(F^{\rm UV}_{\cal S}) &\stackrel{D=6-2\eps}{=}& 
-\frac{\alpha}{\eps} \, 
+ 
\frac{\alpha^2}{\eps} \frac{1}{2} \, 
+
 \frac{\alpha^3}{\eps}\,
\left(\frac{\zeta_3}{3}-\frac{17}{18}\right) \,
  + \cO(\alpha^4,\epsilon^0) \, , \quad {\rm with}\;\;\; \alpha = {-q^2}  \frac{g^2 N}{ (4 \pi)^3}\,.
\end{align}
Let us now discuss this result.

Despite the fact that the form factor is better behaved in the UV than
expected, one may wonder why the four-particle amplitudes at one- and
two loops are even better behaved in the UV than the form factor.
This is due to the fact that there are specific counterterms for the local composite operator ${\cal O}(x)$ in
higher dimensions. Another way of saying this is in terms of operator mixing.
We note that in $D$ dimensions, the coupling constant $g$ has dimension $(4-D)/2$.
Therefore, in $D=6$, the operator ${\rm tr}\,( \phi^2 )$ can mix at one loop with the operator
$g^2 \, \square\, {\rm tr}\, ( \phi^2 )$, and other operators having the same quantum numbers 
(we have dropped $SU(4)$ indices for simplicity).
Another reason for the better UV behaviour of the four-point amplitudes, at least in the planar limit, 
is the fact that amplitudes have a dual conformal symmetry, which implies that
the difference between the number of propagator factors and numerator factors is four for any loop, 
whereas form factors are not dual conformal invariant and therefore can
have fewer propagators per loop.


\section{Discussion and conclusion}\label{sec:conc}

In this paper, we extended the calculation of the two-particle form factor in $\cN=4$ SYM of
ref. \cite{vanNeerven:1985ja} to the three-loop order.
We employed the unitarity-based method to obtain the answer in terms of
loop integrals. The result contains both planar and non-planar integrals.

The form factor can be expressed in several ways in terms of loop integrals that make
different properties manifest. One way of writing it, Eq.~(\ref{eq:3lfirst}) is in terms of 
integrals all having uniform transcendentality (UT). Other forms,
Eqs.~(\ref{eq:FFUV}) and~(\ref{eq:FFUVmanifest}),
do not have this property, but in turn have the advantage of making the ultraviolet
properties of the form factor manifest.
In order to see the connection between the two representations, we derived identities between
non-planar integrals based on reparametrisation invariances.

We evaluated the form factor in dimensional regularisation
by reexpressing the integrals appearing in it in terms of 
conventionally used master integrals, c.f. Eq.~(\ref{eq:app:3lff}), whose $\eps$ expansion
is known. This allowed us to evaluate the form factor to $\cO(\epsilon^2)$. 
We verified the expected exponentiation of infrared divergences,
with the correct values at three loops of the cusp and collinear anomalous dimensions.

We observed that the heuristic leading transcendentality principle that
relates anomalous dimensions in QCD with those in $\cN=4$ SYM 
holds also for the form factor. We checked this principle to three loops,
up to and including terms of transcendental weight eight. 

We also studied the ultraviolet (UV) properties of the form factor in higher dimensions.
We found that at three loops the UV behaviour is better than suggested by a 
supersymmetry argument. Based on power counting one would expect three-loop integrals 
having $8$ propagators (or nine propagators, and one loop-dependent numerator factor) 
to diverge in $D=16/3$ dimensions. However, we find that the particular linear combinations of integrals 
appearing in the form factor is in fact finite in this dimension, and diverges only in $D=6$.
We found a form, Eqs.~(\ref{eq:FFUV}) and~(\ref{eq:FFUVmanifest}), where this is manifest, and computed the leading UV
divergence of $\log(F_{\cal S})$ in $D=6-2 \eps$ dimensions.

There are a number of interesting further directions.

It is interesting to compare the UV behaviour of the form factor to
that of four-particle scattering amplitudes. While there are differences due to 
specific counterterms allowed for composite operators, 
they both share the property of having better UV behaviour than expected.
It would be interesting if one could understand the UV behaviour of the
form factor a priori, perhaps based on the absence of potential counterterms,
or from string theory arguments.

We remark that the representations of the form factor in terms of UT integrals, Eq.~(\ref{eq:3lfirst}),
or those making its ultraviolet properties manifest, Eq.~(\ref{eq:FFUVmanifest}), are simpler than
that in terms of conventionally used master integrals. This may indicate that, even beyond
$\cN=4$ SYM, there exists a basis of integrals in terms of which the result looks simpler.
Similar observations about the simplicity of loop integrands and integrals in the case
of planar scattering amplitudes were also made in refs. \cite{ArkaniHamed:2010kv} and \cite{Drummond:2010mb}.

A further extension of this work could be to investigate 
generalised form factors with more on-shell external legs. At one-loop 
even all-multiplicity results could be 
envisaged~\cite{Brandhuber:2010ad,Brandhuber:2011tv,Bork:2010wf,Bork:2011cj}.
At two loops, at least the three-particle form factors should be 
computable in a relatively straightforward manner, since the 
relevant integrals (two-loop four-point functions with one external 
leg off-shell,~\cite{3jmi}) are known from the calculation of QCD amplitudes for 
the $1\to 3$ decay kinematics~\cite{3jtensor,htensor}.

The form factor studied in this paper has a very rich structure, similar to that
of scattering amplitudes. Planar loop integrands of scattering amplitudes, just like tree amplitudes, 
satisfy powerful recursion relations \cite{ArkaniHamed:2010kv}. 
It would be extremely interesting to extend the applicability of recursion relations to the
non-planar case, and the form factor studied here is perhaps the simplest case of this type 
where non-planar integrals appear. 


\section*{Acknowledgments}
It is a pleasure to thank L.~Dixon for many stimulating discussions and for sharing his insights on
the transcendentality properties of non-planar loop integrals with us.
We would also like to thank Z.~Bern, H.~Johansson, S.~Naculich, R.~Roiban, and E.~Sokatchev for useful discussions.
This project was started during the ``Harmony of Scattering Amplitudes'' program 
at the KITP Santa Barbara, whose  hospitality and support we
gratefully acknowledge. 
This research was 
supported in part by the National Science Foundation under Grant No. PHY05-51164.
The work of TG was supported by the Swiss National Science Foundation 
(SNF) under grant 200020-138206, 
JMH was supported in part by the Department of Energy grant DE-FG02-90ER40542,
TH is supported by the Helmholtz Alliance ``Physics at the Terascale''. 
Diagrams were drawn with axodraw~\cite{Vermaseren:1994je}.


\appendix

\section{Explicit results of integrals}\label{app:integrals}

In this appendix we list explicit expressions of the integrals that
appear as building blocks of the form factor. Our integration measure
per loop reads
\be\label{app:measure}
\dps \int \!\!\frac{d^Dk}{i(2\pi)^D} \; ,
\ee
and we define the pre-factor
\be
\dps S_\Gamma = \frac{1}{(4\pi)^{D/2} \, \Gamma(1-\eps)} \, .
\ee
A generic integral $I$ can be decomposed according to
\be
\dps I = S_\Gamma^L \left[-q^2-i\eta\right]^{n-L\eps}
\cdot I^{{\rm exp}} \; ,
\ee
where $L$ is the number of loops, and the integer $n$ is fixed by
dimensional arguments. $I^{{\rm exp}}$ contains the Laurent expansion
about $\eps=0$.

We start with the one-loop integral
{\allowdisplaybreaks
\bea
D_1 &=& S_\Gamma \left[-q^2-i\eta\right]^{-1-\eps}\cdot D_1^{{\rm
exp}} \, ,\nnb \\
D_1^{{\rm exp}} &=&
-\frac{\Gamma^2(-\eps)\Gamma(1-\eps)\Gamma(1+\eps)}{\Gamma(1-2\eps)}
\, .
\eea
}

At two loops the integrals read
{\allowdisplaybreaks
\bea
E_1 &=& S_\Gamma^2 \left[-q^2-i\eta\right]^{-2-2\eps}\cdot E_1^{{\rm
exp}} \, ,\nnb \\
E_1^{{\rm exp}} &=&\frac{\Gamma^2(1-\eps) \Gamma^2(\eps+1)
\Gamma^4(-\eps)}{\Gamma^2(1-2 \eps)}-\frac{3 \Gamma (1-\eps) \Gamma
(2\eps+1) \Gamma^4(-\eps)}{2 \Gamma (1-3 \eps)} \nnb \\
&&+\frac{3 \Gamma (1-2\eps) \Gamma (\eps+1) \Gamma (2
\eps+1)\Gamma^4(-\eps)}{4 \Gamma (1-3 \eps)} \, .
\eea
}
An all-order expression for $E_2$ can be found
in~\cite{Gehrmann:2005pd}. The expansion in $\eps$ reads
{\allowdisplaybreaks
\bea
E_2 &=& S_\Gamma^2 \left[-q^2-i\eta\right]^{-2-2\eps}\cdot E_2^{{\rm
exp}} \, ,\nnb \\
E_2^{{\rm exp}} &=&+\frac{1}{\eps^4}-\frac{5 \pi ^2}{6 \eps^2}-\frac{27 \zeta_3}{\eps}-\frac{23 \pi ^4}{36}
+\eps \left(8 \pi ^2 \zeta_3-117 \zeta_5\right)
+\eps^2 \left(267 \zeta_3^2-\frac{19 \pi ^6}{315}\right) \nnb \\
&&+\eps^3 \left(\frac{109 \pi ^4 \zeta_3}{10}+40 \pi ^2 \zeta_5+6\zeta_7\right)
+\eps^4 \left(-264 \zeta_{5,3}+2466 \zeta_3 \zeta_5-44 \pi ^2\zeta_3^2+\frac{1073 \pi^8}{3024}\right)\nnb \\
&& + {\cal{O}}(\eps^5) \, .
\eea
}

At three loops the integrals with uniform transcendentality (UT) are
shown in Figs.~\ref{fig:34pcuts} and~\ref{fig:diagsF} and read
{\allowdisplaybreaks
\bea
F_1 &=& S_\Gamma^3 \left[-q^2-i\eta\right]^{-3-3\eps}\cdot F_1^{{\rm
exp}} \, ,\nnb \\
F_1^{{\rm exp}} &=&-\frac{1}{36 \eps^6}-\frac{\pi ^2}{12 \eps^4}-\frac{31 \zeta_3}{18 \eps^3}-\frac{23 \pi ^4}{216 \eps^2}
+\frac{1}{\eps}\left(-\frac{5 \pi ^2 \zeta_3}{6}-\frac{49\zeta_5}{2}\right)\nnb\\
&& -\frac{43 \zeta_3^2}{18}-\frac{5657 \pi ^6}{68040}
+\eps \left(\frac{227 \pi ^4 \zeta_3}{540}-\frac{7 \pi ^2 \zeta_5}{6}-\frac{139 \zeta_7}{3}\right)\nnb\\
&& +\eps^2 \left(-192 \zeta_{5,3}+3 \zeta_3 \zeta_5+\frac{47 \pi ^2 \zeta_3^2}{2}+\frac{959 \pi ^8}{12960}\right)
+ {\cal{O}}(\eps^3) \, .
\eea
}
The integral $F_2$ is just $A_{9,1}^{(n)}$ from~\cite{masterC},
{\allowdisplaybreaks
\bea
F_2 &=& S_\Gamma^3 \left[-q^2-i\eta\right]^{-2-3\eps}\cdot F_2^{{\rm
exp}} \, ,\nnb \\
F_2^{{\rm exp}} &=&+\frac{1}{36 \eps^6}+\frac{\pi ^2}{18 \eps^4}+\frac{14 \zeta_3}{9 \eps^3}+\frac{47 \pi ^4}{405 \eps^2}
+\frac{1}{\eps}\left(\frac{85 \pi ^2 \zeta_3}{27}+20 \zeta_5\right)\nnb \\
&&+\frac{137 \zeta_3^2}{3}+\frac{1160 \pi ^6}{5103} +\eps \left(\frac{829 \pi ^4 \zeta_3}{405}+\frac{719 \pi ^2 \zeta_5}{27}+\frac{6451 \zeta_7}{9}\right)\nnb\\
&&+\eps^2 \left(-\frac{1184}{9} \zeta_{5,3}+1250 \zeta_3
\zeta_5-\frac{712 \pi^2 \zeta_3^2}{9}+\frac{593749 \pi^8}{1224720}\right) + {\cal{O}}(\eps^3) \, .
\eea
}
Moreover, we have
{\allowdisplaybreaks
\bea
F_3 &=& S_\Gamma^3 \left[-q^2-i\eta\right]^{-2-3\eps}\cdot F_3^{{\rm
exp}} \, ,\nnb \\
F_3^{{\rm exp}} &=&-\frac{1}{36 \eps^6}+\frac{\pi ^2}{9 \eps^4}+\frac{37 \zeta_3}{9 \eps^3}+\frac{131 \pi ^4}{540 \eps^2}
+\frac{1}{\eps}\left(\frac{145 \zeta_5}{3}-\frac{4 \pi ^2 \zeta_3}{9}\right)\nnb \\
&&-\frac{1352 \zeta_3^2}{9}+\frac{173 \pi ^6}{1215}+\eps \left(-\frac{253 \pi ^4 \zeta_3}{27}-\frac{62 \pi ^2 \zeta_5}{3}-\frac{525 \zeta_7}{2}\right)\nnb\\
&& +\eps^2 \left(\frac{6272}{5} \zeta_{5,3}-\frac{4696 \zeta_3 \zeta_5}{3}-\frac{712 \pi ^2 \zeta_3^2}{9}-\frac{1301609 \pi ^8}{1701000}\right)+ {\cal{O}}(\eps^3) \, , \\
&&\nnb \\
F_4 &=& S_\Gamma^3 \left[-q^2-i\eta\right]^{-2-3\eps}\cdot F_4^{{\rm
exp}} \, ,\nnb \\
F_4^{{\rm exp}} &=&-\frac{1}{36 \eps^6}-\frac{\pi ^2}{12 \eps^4}-\frac{55 \zeta_3}{18 \eps^3}-\frac{11 \pi ^4}{216 \eps^2}
+\frac{1}{\eps}\left(\frac{43 \pi ^2 \zeta_3}{6}-\frac{599
\zeta_5}{6}\right)\nnb \\
&&-\frac{307 \zeta_3^2}{18}-\frac{18797 \pi ^6}{68040}+\eps \left(-\frac{149 \pi ^4 \zeta_3}{108}+\frac{239 \pi ^2 \zeta_5}{2}-\frac{21253
\zeta_7}{6}\right) \nnb\\
&& +\eps^2 \left(\frac{8268}{5} \zeta_{5,3}+\frac{5569 \zeta_3 \zeta_5}{3}-\frac{439 \pi ^2 \zeta_3^2}{6}-\frac{184873 \pi ^8}{108000}\right)+ {\cal{O}}(\eps^3) \, , \\
&&\nnb \\
F_5 &=& S_\Gamma^3 \left[-q^2-i\eta\right]^{-2-3\eps}\cdot F_5^{{\rm
exp}} \, ,\nnb \\
F_5^{{\rm exp}} &=&+\frac{1}{12 \eps^6}+\frac{\pi ^2}{27 \eps^4}+\frac{17 \zeta_3}{9 \eps^3}+\frac{71 \pi ^4}{540 \eps^2}
+\frac{1}{\eps}\left(\frac{71 \pi ^2 \zeta_3}{54}+\frac{13\zeta_5}{3}\right)\nnb\\
&&-\frac{679 \zeta_3^2}{6}+\frac{3991 \pi ^6}{136080}+\eps \left(-\frac{2837 \pi ^4 \zeta_3}{540}+\frac{205 \pi ^2 \zeta_5}{9}-\frac{25135 \zeta_7}{24}\right)\nnb\\
&& +\eps^2 \left(\frac{4006}{3} \zeta_{5,3}-59 \zeta_3 \zeta_5-\frac{10 \pi ^2 \zeta_3^2}{27}-\frac{14156063 \pi ^8}{16329600}\right)+ {\cal{O}}(\eps^3) \, .
\eea
}
The integral $F_6$ is just $A_{9,2}^{(n)}$ from~\cite{masterC},
{\allowdisplaybreaks
\bea
F_6 &=& S_\Gamma^3 \left[-q^2-i\eta\right]^{-2-3\eps}\cdot F_6^{{\rm
exp}} \, ,\nnb \\
F_6^{{\rm exp}} &=&+\frac{2}{9 \eps^6}-\frac{7 \pi ^2}{27 \eps^4}-\frac{91 \zeta_3}{9 \eps^3}-\frac{373 \pi ^4}{1080 \eps^2}
+\frac{1}{\eps}\left(\frac{179 \pi ^2 \zeta_3}{27}-167 \zeta_5\right)\nnb\\
&&+\frac{169 \zeta_3^2}{9}-\frac{59797 \pi ^6}{136080}+\eps \left(\frac{7 \pi ^4 \zeta_3}{30}+\frac{850 \pi ^2 \zeta_5}{9}-\frac{18569 \zeta_7}{6}\right)\nnb\\
&& + \eps^2 \left(\frac{5188}{5} \zeta_{5,3}+\frac{9362 \zeta_3 \zeta_5}{3}-\frac{4436 \pi ^2 \zeta_3^2}{27}-\frac{107881603 \pi ^8}{81648000}\right) + {\cal{O}}(\eps^3) \, .
\eea
}
Moreover, we have
{\allowdisplaybreaks
\bea
F_7 &=& S_\Gamma^3 \left[-q^2-i\eta\right]^{-1-3\eps}\cdot F_7^{{\rm
exp}} \, ,\nnb \\
F_7^{{\rm exp}} &=&-\frac{1}{36 \eps^6}-\frac{\pi ^2}{27 \eps^4}-\frac{7 \zeta_3}{9 \eps^3}-\frac{\pi ^4}{36 \eps^2}
+\frac{1}{\eps}\left(\frac{20 \pi ^2 \zeta_3}{27}-\frac{13 \zeta_5}{3}\right)\nnb\\
&&+\frac{226 \zeta_3^2}{9}-\frac{233 \pi ^6}{34020}+\eps \left(\frac{151 \pi ^4 \zeta_3}{135}+\frac{70 \pi ^2 \zeta_5}{9}-\frac{229 \zeta_7}{6}\right)\nnb\\
&& +\eps^2 \left(\frac{248}{15} \zeta_{5,3}+\frac{1244 \zeta_3 \zeta_5}{3}-\frac{176 \pi ^2 \zeta_3^2}{27}+\frac{207311 \pi ^8}{20412000}\right)+ {\cal{O}}(\eps^3) \, , \\
&&\nnb \\
F_8 &=& S_\Gamma^3 \left[-q^2-i\eta\right]^{-2-3\eps}\cdot F_8^{{\rm
exp}} \, ,\nnb \\
F_8^{{\rm exp}} &=&+\frac{1}{36 \eps^6}+\frac{\pi ^2}{27 \eps^4}-\frac{5 \zeta_3}{9 \eps^3}+\frac{\pi ^4}{108 \eps^2}
+\frac{1}{\eps}\left(\frac{37 \zeta_5}{3}-\frac{32 \pi ^2 \zeta_3}{27}\right)\nnb\\
&&+\frac{98 \zeta_3^2}{9}+\frac{26 \pi ^6}{8505}+\eps \left(-\frac{4 \pi ^4 \zeta_3}{15}-\frac{70 \pi ^2 \zeta_5}{9}+\frac{835 \zeta_7}{6}\right)\nnb\\
&&+\eps^2 \left(\frac{248}{3} \zeta_{5,3}+\frac{124 \zeta_3 \zeta_5}{3}+\frac{572 \pi ^2 \zeta_3^2}{27}-\frac{16159 \pi ^8}{1020600}\right) + {\cal{O}}(\eps^3) \, , \\
&&\nnb \\
F_9 &=& S_\Gamma^3 \left[-q^2-i\eta\right]^{-2-3\eps}\cdot F_9^{{\rm
exp}} \, ,\nnb \\
F_9^{{\rm exp}} &=&+\frac{1}{4 \eps^6}-\frac{11 \pi ^2}{54 \eps^4}-\frac{74 \zeta_3}{9 \eps^3}-\frac{43 \pi ^4}{180 \eps^2}
-\frac{1}{\eps}\left(\frac{328 \zeta_5}{3}-\frac{176 \pi ^2 \zeta_3}{27}\right)\nnb\\
&&+128 \zeta_3^2-\frac{2951 \pi ^6}{17010}-\eps \left(-\frac{1021 \pi ^4 \zeta_3}{135}-\frac{610 \pi ^2 \zeta_5}{9}+\frac{6149 \zeta_7}{6}\right)\nnb\\
&& -\eps^2 \left(\frac{392}{3} \zeta_{5,3}-\frac{11504 \zeta_3 \zeta_5}{3}+\frac{2876 \pi ^2 \zeta_3^2}{27}-\frac{85171 \pi ^8}{1020600}\right)+ {\cal{O}}(\eps^3) \, , \\
&&\nnb \\
F_{10} &=& S_\Gamma^3 \left[-q^2-i\eta\right]^{-1-3\eps}\cdot F_{10}^{{\rm
exp}} \, ,\nnb \\
F_{10}^{{\rm exp}} &=& \frac{\Gamma (1-\eps)^2 \Gamma (-\eps)^5 \Gamma
(3 \eps)}{12 \Gamma (1-4 \eps)} \, .
\eea
}
{}
\FIGURE[t]{
\includegraphics[width=0.95\textwidth]{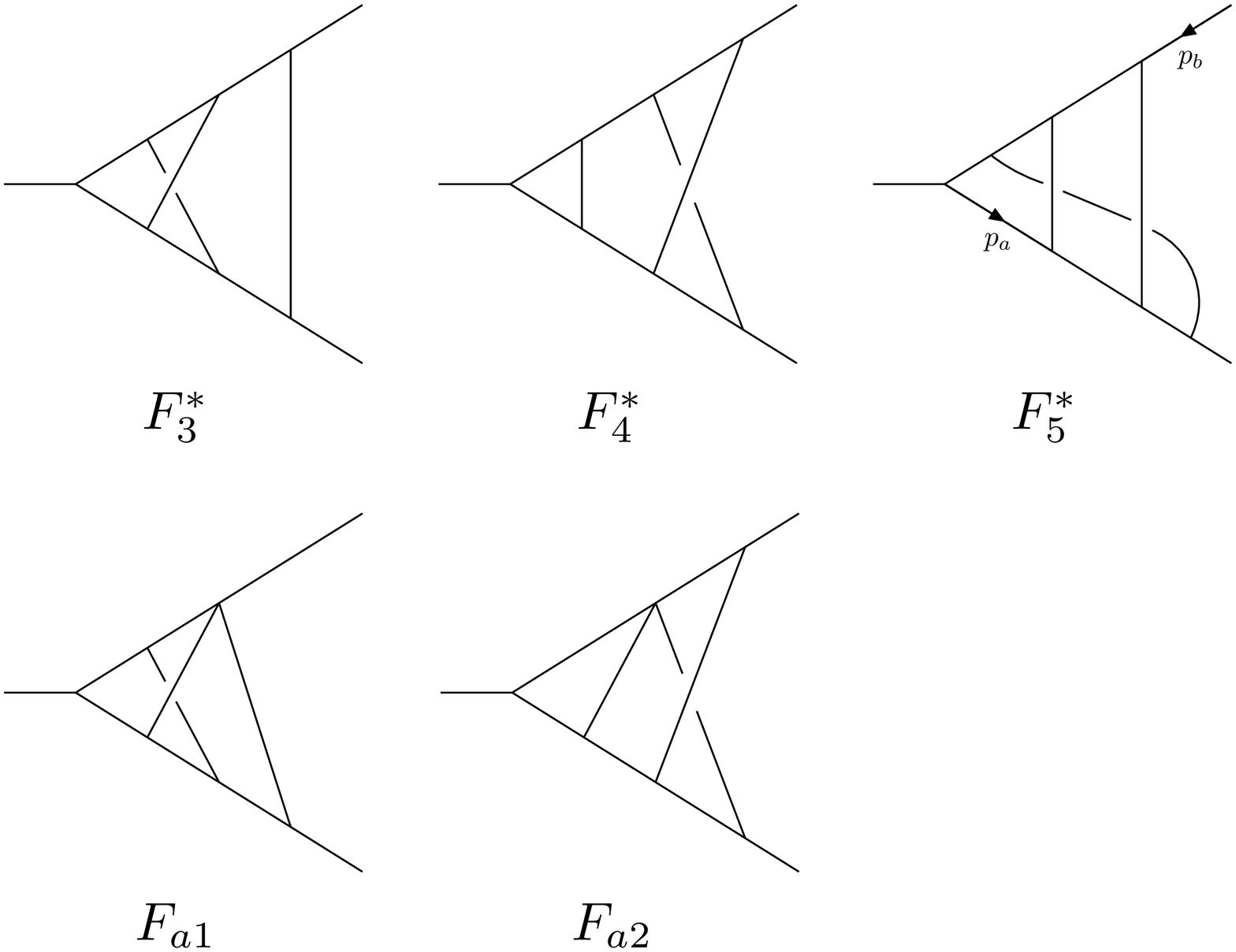}
\caption{Diagrams which do {\textit{not}} have uniform
transcendentality. As before, labels $p_a$ and
$p_b$ on arrow lines indicate an irreducible scalar product
$(p_a+p_b)^2$ in the respective numerator, and diagrams that lack these
labels have unit numerator.}
\label{fig:diagsFstar}}
The integrals {\textit{without}} homogeneous transcendental weight
are collected in Fig.~\ref{fig:diagsFstar} and read
{\allowdisplaybreaks
\bea
F_3^{\ast} &=& S_\Gamma^3 \left[-q^2-i\eta\right]^{-3-3\eps}\cdot
F_3^{\ast \, {\rm exp}} \, ,\nnb \\
F_3^{\ast \, {\rm exp}} &=& -\frac{1}{9 \eps^6}+\frac{4 \pi ^2}{27
\eps^4}+\frac{1}{\eps^3}\left(\frac{28 \zeta_3}{3}+\frac{2 \pi ^2}{9}\right)
+\frac{1}{\eps^2}\left(\frac{44 \zeta_3}{3}-\frac{8 \pi ^2}{9}+\frac{7 \pi ^4}{15}\right) \nnb\\
&& +\frac{1}{\eps}\left(-\frac{176 \zeta_3}{3}+\frac{40 \pi ^2 \zeta_3}{27}+72 \zeta_5+\frac{32 \pi ^2}{9}+\frac{8 \pi ^4}{15}\right)
-\frac{236 \zeta_5}{3}-\frac{2900 \zeta_3^2}{9}+\frac{56 \pi ^2 \zeta_3}{9}\nnb\\
&& +\frac{704 \zeta_3}{3}+\frac{158 \pi ^6}{567}-\frac{32 \pi ^4}{15}-\frac{128 \pi ^2}{9}
+\eps \left(-\frac{2816 \zeta_3}{3}-\frac{224 \pi ^2\zeta_3}{9}-\frac{2458 \pi ^4 \zeta_3}{135}\right.\nnb\\
&& \left.-\frac{1936\zeta_3^2}{3}+\frac{944 \zeta_5}{3}-\frac{232 \pi ^2
\zeta_5}{9}-\frac{2410 \zeta_7}{3}+\frac{512 \pi ^2}{9}+\frac{128 \pi ^4}{15}-\frac{262 \pi ^6}{945}\right) \nnb \\
&&
+\eps^2 \!\left(\frac{35152}{15} \zeta_{5,3}-\frac{16082\zeta_7}{3}-\frac{9640 \zeta_3 \zeta_5}{3}-\frac{352 \pi ^2 \zeta_5}{3}
-\frac{3776 \zeta_5}{3}-\frac{5416 \pi ^2 \zeta_3^2}{27}-\frac{512 \pi ^4}{15}\right.\nnb \\
&&\left.+\frac{7744 \zeta_3^2}{3}-\frac{224 \pi ^4 \zeta_3}{9}+\frac{896 \pi ^2 \zeta_3}{9}
+\frac{11264 \zeta_3}{3}-\frac{956008 \pi ^8}{637875}+\frac{1048 \pi ^6}{945}-\frac{2048 \pi ^2}{9}\right)
\nnb \\
&& + {\cal{O}}(\eps^3) \, , \\
&&\nnb \\
F_4^{\ast} &=& S_\Gamma^3 \left[-q^2-i\eta\right]^{-3-3\eps}\cdot
F_4^{\ast \, {\rm exp}} \, ,\nnb \\
F_4^{\ast \, {\rm exp}} &=& -\frac{1}{18 \eps^6}+\frac{5}{18\eps^5}+\frac{1}{\eps^4}\left(-\frac{10}{9}-\frac{\pi ^2}{6}\right)
+\frac{1}{\eps^3}\left(-\frac{55 \zeta_3}{9}+\frac{40}{9}-\frac{7 \pi ^2}{9}\right)  \nnb\\
&& +\frac{1}{\eps^2}\left(-\frac{136 \zeta_3}{9}-\frac{160}{9}+\frac{28 \pi ^2}{9}-\frac{11 \pi ^4}{108}\right)
+\frac{1}{\eps}\left(\frac{544\zeta_3}{9}+\frac{43\pi^2\zeta_3}{3}-\frac{599\zeta_5}{3}\right.\nnb \\
&& \left.+\frac{640}{9}-\frac{112\pi^2}{9}-\frac{17\pi^4}{54}\right)
-\frac{1108 \zeta_5}{3}-\frac{307 \zeta_3^2}{9}+\frac{88 \pi ^2 \zeta_3}{9}-\frac{2176 \zeta_3}{9}-\frac{18797 \pi ^6}{34020} \nnb \\
&& +\frac{34 \pi ^4}{27}+\frac{448 \pi ^2}{9}-\frac{2560}{9} 
+\eps \left(\frac{8704 \zeta_3}{9}-\frac{352 \pi ^2\zeta_3}{9}-\frac{149 \pi ^4 \zeta_3}{54}
-\frac{7360 \zeta_3^2}{9}+\frac{4432 \zeta_5}{3}\right.\nnb \\
&& \left.+239 \pi ^2 \zeta_5-\frac{21253 \zeta_7}{3}+\frac{10240}{9}-\frac{1792 \pi ^2}{9}-\frac{136 \pi ^4}{27}-\frac{3055 \pi ^6}{1701}\right)
+\eps^2 \left(\frac{16536}{5} \zeta_{5,3}\right. \nnb \\
&& -17273 \zeta_7+\frac{11138\zeta_3 \zeta_5}{3}+180 \pi ^2 \zeta_5-\frac{17728 \zeta_5}{3}-\frac{439 \pi ^2 \zeta_3^2}{3}
+\frac{29440 \zeta_3^2}{9}-\frac{4846 \pi ^4 \zeta_3}{135} \nnb \\
&& \left.+\frac{1408 \pi ^2 \zeta_3}{9}-\frac{34816 \zeta_3}{9}-\frac{184873 \pi ^8}{54000}+\frac{12220 \pi ^6}{1701}+\frac{544 \pi ^4}{27}+\frac{7168 \pi ^2}{9}-\frac{40960}{9}\right) \nnb \\
&& + {\cal{O}}(\eps^3) \, , \\
&&\nnb \\
F_5^{\ast} &=& S_\Gamma^3 \left[-q^2-i\eta\right]^{-2-3\eps}\cdot
F_5^{\ast \, {\rm exp}} \, ,\nnb \\
F_5^{\ast \, {\rm exp}} &=& -\frac{1}{18 \eps^6}-\frac{5}{36 \eps^5}+\frac{1}{\eps^4}\left(\frac{5}{9}+\frac{\pi ^2}{54}\right)+\frac{1}{\eps^3}\left(\frac{5 \pi ^2}{18}-\frac{20}{9}\right)  \nnb\\
&& +\frac{1}{\eps^2}\left(\frac{2 \zeta_3}{9}+\frac{80}{9}-\frac{10 \pi ^2}{9}-\frac{\pi ^4}{40}\right)
+\frac{1}{\eps}\left(-\frac{8 \zeta_3}{9}-\frac{77 \pi ^2 \zeta_3}{54}+\frac{160 \zeta_5}{3}\right.\nnb \\
&& \left.-\frac{320}{9}+\frac{40 \pi ^2}{9}-\frac{59 \pi ^4}{540}\right)+224 \zeta_5+\frac{4003 \zeta_3^2}{18}-8 \pi ^2 \zeta_3+\frac{32 \zeta_3}{9}+\frac{16099 \pi ^6}{68040} \nnb \\
&& +\frac{59 \pi ^4}{135}-\frac{160 \pi ^2}{9}+\frac{1280}{9}
+\eps \left(-\frac{128 \zeta_3}{9}+32 \pi ^2 \zeta_3+\frac{151 \pi ^4 \zeta_3}{12}+\frac{6584 \zeta_3^2}{9}-896 \zeta_5\right.\nnb \\
&& \left. -\frac{445 \pi ^2 \zeta_5}{9}+\frac{74815 \zeta_7}{24}-\frac{5120}{9}+\frac{640 \pi ^2}{9}-\frac{236 \pi ^4}{135}+\frac{2519 \pi ^6}{2430}\right)
+\eps^2 \left(-\frac{12518}{5} \zeta_{5,3}\right.\nnb \\
&& +\frac{67901 \zeta_7}{6}+773 \zeta_3 \zeta_5-\frac{94 \pi ^2 \zeta_5}{3}+3584 \zeta_5+\frac{1570 \pi ^2 \zeta_3^2}{27}-\frac{26336 \zeta_3^2}{9}+\frac{4103 \pi ^4 \zeta_3}{135} \nnb \\
&& \left.-128 \pi ^2 \zeta_3+\frac{512 \zeta_3}{9}+\frac{13248257 \pi ^8}{5832000}-\frac{5038 \pi ^6}{1215}+\frac{944 \pi ^4}{135}-\frac{2560 \pi ^2}{9}+\frac{20480}{9}\right) \nnb \\
&& + {\cal{O}}(\eps^3) \, , \\
&&\nnb \\
F_{a1} &=& S_\Gamma^3 \left[-q^2-i\eta\right]^{-2-3\eps}\cdot
F_{a1}^{{\rm exp}} \, ,\nnb \\
F_{a1}^{{\rm exp}} &=& -\frac{\pi ^2}{9 \eps^3} +\frac{1}{\eps^2}\left(\frac{4 \pi ^2}{9}-\frac{22 \zeta_3}{3}\right)
+\frac{1}{\eps}\left(\frac{88 \zeta_3}{3}-\frac{16 \pi ^2}{9}-\frac{4 \pi ^4}{15}\right)+\frac{118 \zeta_5}{3}-\frac{28 \pi ^2 \zeta_3}{9}\nnb\\
&& -\frac{352 \zeta_3}{3}+\frac{16 \pi ^4}{15}+\frac{64 \pi ^2}{9} 
+\eps \left(\frac{1408 \zeta_3}{3}+\frac{112 \pi ^2 \zeta_3}{9}+\frac{968 \zeta_3^2}{3}-\frac{472 \zeta_5}{3}-\frac{256 \pi ^2}{9}\right.\nnb \\
&& \left.-\frac{64 \pi ^4}{15}+\frac{131 \pi ^6}{945}\right)
+\eps^2 \left(-\frac{5632 \zeta_3}{3}-\frac{448 \pi ^2 \zeta_3}{9}+\frac{112 \pi ^4 \zeta_3}{9}-\frac{3872 \zeta_3^2}{3}+\frac{1888 \zeta_5}{3}\right.\nnb \\
&&\left.+\frac{176 \pi ^2 \zeta_5}{3}+\frac{8041 \zeta_7}{3}+\frac{1024 \pi ^2}{9}+\frac{256 \pi ^4}{15}-\frac{524 \pi ^6}{945}\right) + {\cal{O}}(\eps^3) \, , \\
&&\nnb \\
F_{a2} &=& S_\Gamma^3 \left[-q^2-i\eta\right]^{-2-3\eps}\cdot
F_{a2}^{{\rm exp}} \, ,\nnb \\
F_{a2}^{{\rm exp}} &=& -\frac{5}{36 \eps^5}+\frac{5}{9 \eps^4}+\frac{1}{\eps^3}\left(\frac{7 \pi ^2}{18}-\frac{20}{9}\right)
+\frac{1}{\eps^2}\left(\frac{68 \zeta_3}{9}+\frac{80}{9}-\frac{14 \pi ^2}{9}\right)
+\frac{1}{\eps}\left(-\frac{272 \zeta_3}{9}-\frac{320}{9}\right. \nnb\\
&& \left.+\frac{56 \pi ^2}{9}+\frac{17 \pi ^4}{108}\right)+\frac{554 \zeta_5}{3}-\frac{44 \pi ^2 \zeta_3}{9}+\frac{1088 \zeta_3}{9}-\frac{17 \pi ^4}{27}-\frac{224 \pi ^2}{9}+\frac{1280}{9}\nnb \\
&& +\eps \left(\frac{176 \pi ^2 \zeta_3}{9}-\frac{4352 \zeta_3}{9}+\frac{3680 \zeta_3^2}{9}-\frac{2216 \zeta_5}{3}-\frac{5120}{9}
+\frac{896 \pi ^2}{9}+\frac{68 \pi ^4}{27}+\frac{3055 \pi ^6}{3402}\right) \nnb \\
&& +\eps^2 \left(\frac{17408 \zeta_3}{9}-\frac{704 \pi ^2 \zeta_3}{9}+\frac{2423 \pi ^4 \zeta_3}{135}-\frac{14720 \zeta_3^2}{9}+\frac{8864 \zeta_5}{3}-90 \pi ^2 \zeta_5+\frac{17273 \zeta_7}{2}\right.\nnb\\
&&\left.+\frac{20480}{9}-\frac{3584 \pi ^2}{9}-\frac{272 \pi ^4}{27}-\frac{6110 \pi ^6}{1701}\right) + {\cal{O}}(\eps^3) \, .
\eea
}
At three loops we also cross-checked the major part of the integrals
with the sector decomposition program
FIESTA~\cite{Smirnov:2008py,Smirnov:2009pb}.


\section{Form factor in terms of master integrals}\label{app:ffmasters}

Just as in QCD, the three-loop scalar form factor in $\mathcal{N}=4$
can be reduced to master integrals by means of the Laporta
algorithm~\cite{Laporta:2001dd}, for which we used the program
REDUZE~\cite{Studerus:2009ye}. One obtains

{\allowdisplaybreaks
\bea
{F}_S^{(3)} &=& R_{\eps}^3 \left[+\frac{(3 D -14)^2}{(D -4) (5 D -22)} \, A_{9,1}\right.
-\frac{2 (3 D -14)}{5 D -22} \, A_{9,2}
-\frac{4 (2 D -9) (3 D -14)}{(D-4) (5 D -22)} \, A_{8,1} \nnb \\
&&-\frac{20 (3 D -13) (D -3)}{(D -4) (2 D -9)} \, A_{7,1}
-\frac{40 (D -3)}{D -4} \, A_{7,2}
+\frac{8 (D -4)}{(2 D -9) (5 D -22)} \, A_{7,3}\nnb \\
&& -\frac{16 (3 D -13) (3 D -11)}{(2 D -9) (5 D -22)} \, A_{7,4}
-\frac{16 (3 D -13) (3 D -11)}{(2 D -9) (5 D -22)} \, A_{7,5} \nnb \\
&& -\frac{128 (2 D -7) (D -3)^2}{3 (D -4) (3 D -14) (5 D -22)} \, A_{6,1} \nnb \\
&&-\frac{16 (2 D -7) (5 D -18) \left(52 D ^2-485 D +1128\right)}{9 (D -4)^2 (2 D -9) (5 D -22)} \, A_{6,2} \nnb \\
&&-\frac{16 (2 D -7) (3 D -14) (3 D -10) (D -3)}{(D -4)^3 (5 D -22)} \, A_{6,3} \nnb \\
&&-\frac{128 (2 D -7) (3 D -8) \left(91 D ^2-821 D +1851\right) (D -3)^2}{3 (D -4)^4 (2 D -9) (5 D -22)} \, A_{5,1} \nnb \\
&&-\frac{128 (2 D -7) \left(1497 D ^3-20423 D ^2+92824 D -140556\right) (D -3)^3}{9 (D -4)^4 (2 D -9) (3 D -14) (5 D -22)} \, A_{5,2} \nnb \\
&&+\frac{4 (D -3)}{D -4} \, B_{8,1}
+\frac{64 (D -3)^3}{(D -4)^3} B_{6,1}
+\frac{48 (3 D -10) (D -3)^2}{(D -4)^3} \, B_{6,2} \nnb \\
&& -\frac{16 (3 D -10) (3 D -8) \left(144 D ^2-1285 D +2866\right) (D -3)^2}{(D -4)^4 (2 D -9) (5 D -22)} \, B_{5,1} \nnb \\
&&+\frac{128 (2 D -7) \left(177 D ^2-1584 D +3542\right) (D -3)^3}{3 (D -4)^4 (2 D -9) (5 D -22)} \, B_{5,2} \nnb \\
&&+\frac{64 (2 D -5) (3 D -8)  (D -3)}{9 (D -4)^5 (2 D -9) (3 D -14) (5 D -22)} \nnb \\
&& \hspace*{10pt} \times \, \left(2502 D ^5-51273 D ^4+419539 D ^3-1713688 D ^2+3495112 D -2848104\right) \, B_{4,1} \nnb \\
&& \left.+\frac{4 (D -3)}{D -4} \, C_{8,1}+\frac{48 (3 D -10) (D -3)^2}{(D -4)^3} \, C_{6,1}\right] \; .
\label{eq:app:3lff}
\eea
}
$R_{\eps}$ is given in Eq.~(\ref{eq:Reps}). In order to arrive at
Eq.~(\ref{eq:3lff}) we have to plug in $D=4-2\eps$ and the
$\eps$-expansions for the master integrals from Eqs.~(A.7) --~(A.27)
of~\cite{Gehrmann:2010ue}, together with their higher order
$\eps$-terms from~\cite{Lee:2010ik}.


\section{Four-point amplitude to two loops}\label{app:4pt}

Here we summarise the known four-point amplitude in $\cN=4$ super Yang-Mills to two loop order.
As we have seen in the main text, both leading and subleading terms in colour are required
when computing the form factor at leading colour using unitarity.

We consider four-point amplitudes in $SU(N)$ gauge theories with all
particles in the adjoint representation. Let us review the decomposition
of the latter into a trace basis with partial amplitudes as coefficients
\cite{FERMILAB-PUB-90-115-T,hep-ph/0201161}. 

At tree-level, we have
\be\label{app:4pttree}
\cA_{4}^{tree} = g^2 \mu^{2 \eps} \sum_{ \sigma \in S_{4}/Z_{4}} \Tr(T^{a_{\sigma(1)}}  T^{a_{\sigma(2)} }T^{a_{\sigma(3)} }T^{a_{\sigma(4)}} ) A_{4;1;1}^{tree}(\sigma(1), \sigma(2),\sigma(3),\sigma(4)) \,,
\ee
where sum goes over the six non-cyclic permutations of $(1234)$,
i.e.\ $ S_{4}/Z_{4} = \{  (1234) ,\allowbreak (2134), \allowbreak (1243), \allowbreak(2314), \allowbreak(3241), \allowbreak(3214) \}$.
The $A_{4;1;1}^{tree}$ are `partial amplitudes'. 
The arguments of $\mathcal{A}$ and $A$ in Eq.~(\ref{app:4pttree}) are abbreviations, i.e.\ $1$ stands for a given particle (gluon,
fermion, or scalar) of a given helicity and momentum $p_{1}^{\mu}$. 
The $T^{a}$ are the $(N^2-1)$ matrices in the fundamental representation of $SU(N)$. 

At loop level, double trace terms are present as well.
Other possible trace terms vanish since ${\rm Tr} (T^{a})=0$ for $SU(N)$. 
We have, at one loop
\bea
\label{app:4pt1loop}
\cA_{4}^{1-loop} &=& g^4 \mu^{4 \eps}  \sum_{ \sigma \in S_{4}/Z_{4}}   N \, \Tr(T^{a_{\sigma(1)}}  T^{a_{\sigma(2)} }T^{a_{\sigma(3)} }T^{a_{\sigma(4)}} ) A_{4;1,1}^{1-loop}(\sigma(1), \sigma(2),\sigma(3),\sigma(4))  \nonumber  \\
&& \hspace{-1.3cm}+ g^4 \mu^{4 \eps} \sum_{ \sigma \in S_{4}/Z_{2}^3}   \Tr(T^{a_{\sigma(1)}}  T^{a_{\sigma(2)} }) \Tr( T^{a_{\sigma(3)} }T^{a_{\sigma(4)}} ) A_{4;1,3}^{1-loop}(\sigma(1), \sigma(2),\sigma(3),\sigma(4)) \,,
\eea
 and two loops \cite{hep-ph/9702424}, 
\bea\label{app:4pt2loop}
\cA_{4}^{2-loop} &=& g^6 \mu^{6 \eps} \sum_{ \sigma \in S_{4}/Z_{4}} \Tr(T^{a_{\sigma(1)}}  T^{a_{\sigma(2)} }T^{a_{\sigma(3)} }T^{a_{\sigma(4)}} ) \times   \\
&& \times  \Big(  N^2 \, A_{4;1,1}^{2-loop,LC}(\sigma(1), \sigma(2),\sigma(3),\sigma(4))  + A_{4;1,1}^{2-loop,SC}(\sigma(1), \sigma(2),\sigma(3),\sigma(4))   \Big)  \nonumber \\
&& + g^6 \mu^{6 \eps}\sum_{ \sigma \in S_{4}/Z_{2}^3}  N \, \Tr(T^{a_{\sigma(1)}}  T^{a_{\sigma(2)} }) \Tr( T^{a_{\sigma(3)} }T^{a_{\sigma(4)}} ) A_{4;1,3}^{2-loop}(\sigma(1), \sigma(2),\sigma(3),\sigma(4)) \,.
\nonumber
\eea
Here $S_{4}/Z_{2}^3 = \{(1234),(1324),(1423)\}$.
The double trace terms are subleading in the expansion in powers of $N$.
At the two-loop order, we also have the appearence of subleading-in-$N$ terms in the single trace terms,
denoted by the superscript $SC$, while the leading-in-$N$ terms have superscript $LC$.

\FIGURE[t]{
\includegraphics[width=1.01\textwidth]{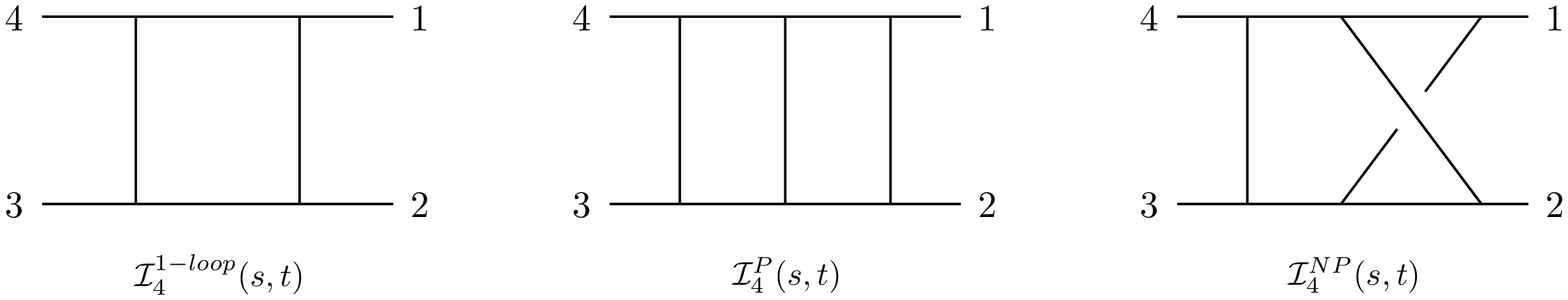}
\caption{Scalar box integrals appearing in four-particle amplitudes to two loops}
\label{fig:boxes}}

$\cN=4$ supersymmetric Ward identities imply that for MHV amplitudes the loop-level
amplitudes are proportional to the tree-level ones, for any choice of external
particles and helicities. We have
\be
A_{4;1,1}^{1-loop}(1,2,3,4)  = - s t  \, A_{4;1,1}^{tree}(1,2,3,4) \, \mathcal{I}_{4}^{1-loop}(s,t) \,,
\ee
where\footnote{Note that our convention of defining loop integrals differs from that
of ref. \cite{hep-ph/9702424} by a factor of $i$ per loop order, cf. Eq.~(\ref{app:measure}).} 
\be
 \mathcal{I}_{4}^{1-loop}(s,t) = \int \frac{d^{D}k}{i (2 \pi)^D}\frac{1}{k^2 (k-p_1 )^2 (k-p_1 - p_2 )^2 (k+p_4 )^2 }\,,
\ee
is the one-loop scalar box integral, see Fig.~\ref{fig:boxes}.
The remaining subleading colour amplitudes at one loop are all equal and given by 
\be
A_{4;1,3}^{1-loop} = \sum_{\sigma \in S_{4}/Z_{4}}  A_{4;1,1}^{1-loop}(\sigma(1), \sigma(2),\sigma(3),\sigma(4)) \,,
\ee
which is the consequence of a $U(1)$ decoupling identity \cite{FERMILAB-PUB-90-115-T}. 

At two loops, the partial amplitudes leading in $N$ are given by \cite{hep-ph/9702424}
\be
A_{4;1,1}^{2-loop,LC}(1,2,3,4)  = + s t  \, A_{4;1,1}^{tree}(1,2,3,4) \, 
\left( 
s \mathcal{I}_{4}^{P}(s,t) +st \mathcal{I}_{4}^{P}(t,s) 
\right)
\,,
\ee 
where $\mathcal{I}_{4}^{P}(s,t)$ is the planar double box integral, see Fig.~\ref{fig:boxes}.

The partial amplitudes subleading in $N$ are given by  \cite{hep-ph/9702424}
\begin{align}
A_{4;1,1}^{2-loop,SC}(1,2,3,4) &= 
2 A_4^{P}(1,2;3,4) + 2 A_4^{P}(3,4;2,1)
 + 2 A_4^{P}(1,4;2,3) + 2 A_4^{P}(2,3;4,1)  \nonumber \\
& \hspace{-2.5 cm} - 4 A_4^{P}(1,3;2,4) - 4 A_4^{P}(2,4;3,1) + 2 A_4^{NP}(1;2;3,4) + 2 A_4^{NP}(3;4;2,1)  \nonumber \\
& \hspace{-2.5 cm} 
+ 2 A_4^{NP}(1;4;2,3) + 2 A_4^{NP}(2;3;4,1) - 4 A_4^{NP}(1;3;2,4) - 4 A_4^{NP}(2;4;3,1) \,,    
\end{align}
and
\begin{align}
A_{4;1,3}^{2-loop}(1;2;3,4) &= 6 A_4^{P}(1,2;3,4) + 6 A_4^{P}(1,2;4,3) + 4 A_4^{NP}(1;2;3,4)    + 4 A_4^{NP}(3;4;2,1) \nonumber  \\
& \hspace{-2 cm} 
- 2 A_4^{NP}(1;4;2,3) - 2 A_4^{NP}(2;3;4,1)
- 2 A_4^{NP}(1;3;2,4) - 2 A_4^{NP}(2;4;3,1) \,, 
\end{align}
where 
\begin{align}
A_4^{P}(1,2;3,4) & \equiv  s_{12}^2 s_{23} \, A_{4;1,1}^{tree}(1,2,3,4)
        \, \mathcal{I}_4^{P}(s_{12}, s_{23}) \,, \\
A_4^{NP}(1;2;3,4) & \equiv  s_{12}^2 s_{23} \,A_{4;1,1}^{tree}(1,2,3,4)
        \, \mathcal{I}_4^{NP}(s_{12}, s_{23}) \,, 
\end{align}
and where $\mathcal{I}_4^{P}$ and $\mathcal{I}_4^{NP}$ are the planar and non-planar double box integral, respectively, see Fig.~\ref{fig:boxes}.

We remark that the expression for the double trace terms $A_{4;1,3}^{2-loop}$ can be obtained 
from the single trace terms using identities derived from group theory \cite{hep-ph/0201161,Naculich:2011ep}.

The tree-level amplitude we need has external scalars only. It is given by
\be
A_{4;1,1}^{tree}(\phi_{12}(1),\phi_{12}(2),\phi_{34}(3),\phi_{34}(4))  = - i\, \frac{s_{12}}{s_{23}} \,.
\ee

\end{document}